\documentclass[sn-mathphys-num]{sn-jnl}

\usepackage{graphicx}%
\usepackage{multirow}%
\usepackage{amsmath,amssymb,amsfonts}%
\usepackage{amsthm}%
\usepackage{mathrsfs}%
\usepackage[title]{appendix}%
\usepackage{xcolor}%
\usepackage{textcomp}%
\usepackage{manyfoot}%
\usepackage{booktabs}%
\usepackage{listings}%
\usepackage{fancyhdr}
\usepackage{lscape}
\usepackage{booktabs}
\usepackage{adjustbox}

\usepackage{array}
\usepackage{longtable}
\usepackage{tabularx}
\usepackage{makecell}
\usepackage{geometry}
\usepackage{caption}

\theoremstyle{thmstyleone}%
\theoremstyle{thmstyletwo}%

\theoremstyle{thmstylethree}%
\begin{document}

\title[Article Title]{Evaluating link prediction: New perspectives and recommendations}


\author[1]{\fnm{Bhargavi Kalyani} \sur{I}} \email{bhargavikalyani\_i@srmap.edu.in}

\author[2]{\fnm{A Rama Prasad} \sur{Mathi}}\email{ramaprasad\_mathi@srmap.edu.in}

\author*[3]{\fnm{Niladri} \sur{Sett}}\email{settniladri@gmail.com}

\affil[1]{\orgdiv{ICT, Parts and Services}, \orgname{Stellantis}, \orgaddress{\street{Gachibowli}, \city{Hyderabad}, \postcode{500032}, \state{Telangana}, \country{India}}}

\affil[2,3]{\orgdiv{Department of Computer Science and Engineering}, \orgname{SRM University AP}, \orgaddress{\city{Neerukonda}, \postcode{522240}, \state{Andhra Pradesh}, \country{India}}}



\abstract{Link prediction (LP) is an important problem in network science and machine learning research. The state-of-the-art LP methods are usually evaluated in a uniform setup, ignoring several factors associated with the data and application specific needs. We identify a number of such factors, such as, network-type, problem-type, geodesic distance between the end nodes and its distribution over the classes, nature and applicability of LP methods, class imbalance and its impact on early retrieval, evaluation metric, etc., and present an experimental setup which allows us to evaluate LP methods in a rigorous and controlled manner. We perform extensive experiments with a variety of LP methods over real network datasets in this controlled setup, and gather valuable insights on the interactions of these factors with the performance of LP through an array of carefully designed hypotheses. Following the insights, we provide recommendations to be followed as best practice for evaluating LP methods.}




\maketitle
\thispagestyle{fancy}
\section{Introduction}
Complex networks encode rich and non-trivial relations among entities. They are ubiquitous in nature, and are used to model many complex real-world systems. Social, biological and information systems are to name a few such systems. Social networks encode some kind of social relationships, biological networks model interactions between two biological entities like proteins, and information networks record information exchanges, such as emails. Each entity in a complex network is called a \textit{node} and any connection, interaction, or exchange of information between two nodes are referred as a \textit{link}. Graph data structure is used to encode complex networks, where vertices represent nodes and edges represent links. \textit{Link prediction (LP)}~\cite{liben2007link,kumar2020link,lichtenwalter2010new} is a fundamental problem in network science, which, given a network, aims at predicting unobserved, missing, hidden or future relationships, depending on the complex system and the target application~\cite{daud2020applications}. LP has numerous applications: it can suggest unobserved or missing relations between two criminals in a criminal network~\cite{calderoni2020robust,berlusconi2016link}, it can predict future collaborations among scientists from co-authorship networks~\cite{sett2018temporal,sun2011co,chuan2018link}, it can recommend unobserved costumer-product relationships in collaborative filtering~\cite{huang2005link,yilmaz2023link,lee2021collaborative}, and so on. LP has been primarily targeted as a classification problem~\cite{kumar2020link} which, given a network, classify a given node pair, which are not connected by a link in the given network, to \textit{yes} or \textit{no}: \textit{yes} signifying they have a hidden connection or they will be connected in the future, and \textit{no} means there is no possibility of getting connected by a link.

Since Nowell et al.~\cite{liben2007link} popularized LP as a research problem, now it is a well established research domain in both of network science and the machine learning community. Application of LP is diverse, so a solution to this problem demands considerations of the application or the domain specific factors. Although, there exists application specific designs of the LP methods~\cite{wang2014link,vahidi2021hybrid,leroy2010cold,yilmaz2023link,li2023effective,sett2018temporal}, they usually extend the state-of-the-art node similarity~\cite{liben2007link,kumar2020link,zhou2009predicting} or machine learning based~\cite{zhang2018link,perozzi2014deepwalk,grover2016node2vec,hamilton2017inductive,he2020lightgcn,menon2011link,yilmaz2023link} methods. These state-of-the-art methods assume a uniform set-up accross different prediction cenerios, starting from problem formulation to experimental setup and evaluation. We investigate if this uniformity hinders accurate evaluation of the LP methods, and aim to provide a guideline for a robust evaluation of LP. We identify a few aspects concerning the data, experimental setup and evaluation metric that may affect the prediction performance, based on which we build our experimental framework, and perform extensive experiments to gain insights toward effective and robust evaluation for LP. We brief those aspects below.

The LP problem can be categorized into two: (a) missing/unobserved/hidden link prediction, and (b) future link prediction~\cite{liben2007link,kumar2020link,martinez2016survey}. Without loss of generality, now onward we refer these two types as \textit{missing LP} and \textit{future LP} respectively, and name this classification as \textit{prediction type}. Traditionally, in literature, the experimental setups for LP do not differentiate between these two prediction types, and evaluate LP in a common experimental setup which ignores the information of the link appearance time, and the  ground truth node pairs are generated by removing a fraction of random edges from the underlying graph. This results in information loss and makes the future LP problem get evaluated as a missing LP problem. This may not capture true efficacy of a prediction method on future LP, because a randomly deleted link may not be a true future link. Our experimental setup accommodates both types of LP, and we examine if this information loss results in any performance degradation of the LP methods. A line of works target \textit{temporal link prediction}~\cite{sett2018temporal,qin2023temporal}, which devices specialized LP methods by exploiting temporal dynamics in dynamic networks. In this work, we are not focusing on temporal link prediction, but consider the traditional approach of solving LP.

Popular LP methods are of two types: \textit{similarity based} and \textit{machine learning based}~\cite{martinez2016survey}. Similarity based methods~\cite{liben2007link} calculate some score based on structural similarity or closeness of a node pair, which measures the likelihood of having a link. These methods can be either \textit{local} or~\textit{global}. Local methods rely on \textit{triadic closure property}~\cite{granovetter1973strength} which says that two unconnected nodes tend to form a link if they have at least one common connection. Local methods are applicable on a pair of nodes whose shortest path distance in the network graph, which is also referred as \textit{geodesic distance} in network science, is two. To increase readability, we refer ``geodesic distance'' as ``distance'' throughout this paper. We refer a node pair having distance two as \textit{two-hop} away/apart node pairs. Global methods are mostly shortest path or random walk based, which can predict connections between any node pair in the network. Modern machine learning based LP methods either train a neural network in a supervised way~\cite{zhang2018link,cai2020multi,pan2021neural,tan2023bring}, or generate low dimension embedding for the nodes in unsupervised manner and use the embeddings for predicting links~\cite{perozzi2014deepwalk,grover2016node2vec,hamilton2017inductive,he2020lightgcn}. Machine learning based methods are global in nature. The \textit{small world property}~\cite{strogatz2001exploring} makes the \textit{diameter} of a complex network grow at a logarithmic rate with network size, and the likelihood of connection decreases with the distance between two nodes. The LP methods which are global in nature, mostly agree with this principal. So, in one hand, the local methods apply on two-hop away node pairs, whereas the global methods apply on two-hop as well as more distant node pairs; on the other hand, the new connection build-up process is biased toward the two-hop away node pairs than the rest. Traditional evaluation frameworks do not distinguish the test node pairs by the distance of the end nodes in the network. So, the evaluation does not capture the unique characteristics, if any, of the test node pairs based on the distance of the end nodes. Moreover, when we compare a global method with a local one in this framework, we may draw wrong conclusions because the test set for the two cases may differ. Also, we miss the opportunity to compare two global methods in a distance controlled setup. The traditional evaluation approach has another inherent issue. The negative test examples are sampled by selecting random node pairs which are not connected by a link in the network, and are not in the positive examples. This randomness makes the distribution of negative examples differ with the positive ones, for which the samples have an inherent distance bias. We propose an experimental set-up which takes care of these factors, and conduct an array of systematic experiments to find valuable insights toward effective evaluation.

The state-of-the-art LP methods usually apply on undirected networks. However, networks like information networks are directed in nature. So, these networks are usually converted to undirected networks by ignoring their directions in order to make those methods applicable on them. Ignoring directions results in loss of significant information about the links, which may affect the prediction performance in directed networks. We investigate this with experiments in our hop controlled and prediction type controlled setup.

Class imbalance is an inherent issue with LP, where the real missing or future links are too few as compared to the node pairs with no possible connections. Historically, \textit{\textbf{A}rea \textbf{U}nder the \textbf{R}eceiver \textbf{O}perating \textbf{C}haracteristics Curve (AUROC)}~\cite{hanley1982meaning} has been a popular single point summary statistics evaluation metric for LP~\cite{kumar2020link,martinez2016survey}, perhaps due to its invariance with class distribution. AUROC favors accurate classification of positive links, at the cost of significantly misclassifying negative node pairs. Hence, AUROC can overestimate the performance of LP. Skew sensitive measures like \textit{\textbf{A}rea \textbf{U}nder the \textbf{P}recision-\textbf{R}ecall Curve (AUPR)} or \textit{Precision@k} can be an alternative~\cite{saito2015precision,lichtenwalter2010new,kumar2020link}, but use of these are rare in literature, may be due to the difficulty in handling the variability in results with skew level for their skew sensitiveness. So, in order to understand how an LP method behaves in skewed test environment, we require a rigorous skew controlled evaluation with appropriate evaluation metrics. Moreover, applications like recommendation systems require methods to perform good at the early retrieval stage. Early retrieval performance is skew sensitive, the effect of which can not be understood with AUROC, or the ROC curve~\cite{truchon2007evaluating,swamidass2010croc,saito2015precision}. So, how the LP methods behave in the early retrieval stage in imbalanced setup has been largely unexplored. Our proposed experiment setup and the conducted experiments take care of these factors and judiciously employ the evaluation metrics to understand the true behavior of the prediction methods considering class imbalance and early retrieval. It also examines the interplay of the other factors like prediction type and distance between the end nodes of the test node pairs with class imbalance.

Our contributions in this paper can be summarized as follows.
\begin{itemize}
 \item we identify a number of aspects concerning the data, experimental setup and evaluation metric that may affect the performance of LP methods, which the baseline setups for the LP evaluation largely ignore;
 \item we propose an experimental setup which allows us to evaluate LP methods in a rigorous manner in a controlled environment, considering those aspects;
 \item we conduct extensive experiments using 24 real network datasets on 58 LP methods in this controlled experimental setup, and perform an array of correlation and hypothesis tests to analyze the effect of those factors on LP methods;
 \item we provide recommendations to be followed as best practice toward an effective and robust evaluation of LP methods.
\end{itemize}

This paper is organized as follows. In Section~\ref{Rel}, we summarize the related works. We provide the details of the LP methods which we use in our experiments in Section~\ref{LP-methods}, grouped by their categories. Section~\ref{datasets} provides the details of the datasets on which we apply the LP methods. We describe the proposed experimental setup in Section~\ref{exp-setup}. In Section \ref{Res_Dis}, we present the results and analyze our findings, and provide our recommendations. We conclude our work and indicate the future directions in Section~\ref{Concl}.

\section{Related Works}
\label{Rel}
Nowell et al.~\cite{liben2007link} proposed a list of local and global similarity based LP methods. Similarity based methods calculate a score between two nodes to estimate their chance to form a link. Local similarity based methods are neighborhood based, and apply to two-hop away node pairs. The global similarity based methods are mostly path based and apply to any node pair in the network. In this paper, we consider most of the methods proposed in Nowell et al. in our experiments along with another popular local method: Resource Allocation~\cite{zhou2009predicting}. There are a few other similarity based methods in literature, an exhaustive list of which can be found in Kumar et al.~\cite{kumar2020link}. We do not include another type of similarity based method, namely quasi-local~\cite{kumar2020link}, in our study. Modern machine learning based LP methods are either supervised~\cite{zhang2018link,cai2020multi,pan2021neural,tan2023bring} or unsupervised~\cite{perozzi2014deepwalk,grover2016node2vec,hamilton2017inductive,he2020lightgcn}. The supervised methods train a neural network or a graph neural network in an end-to-end manner with the test node pairs. The unsupervised methods first train a neural network or a graph neural network to learn low dimension embeddings of the nodes in an unsupervised manner (known as contrastive learning), and then produce link features which are used to build a classifier to predict links. In this paper, we consider the unsupervised methods in our experiments. The main reason to exclude the end-to-end methods is the difficulty in building the experimental set-up for those in the longitudinal setup for the future LP, which incurs a longitudinal bias~\cite{lichtenwalter2010new,yang2015evaluating} on the prediction performance, as well as makes the methods incomparable with their missing LP counterpart. Another group of unsupervised LP approaches are matrix factorization based~\cite{menon2011link,chen2017link}, which we do not include in our experiments. Note that, the purpose of this work is not to present an exhaustive performance comparison of the existing methods, but to devise a robust evaluation strategy for the prediction methods considering certain data,  evaluation metric, experimental setup, application and methodology specific factors. So, excluding a few prediction methods does not hurt the merit of this work.

The start-of-the-art LP methods briefed above mainly apply to a simple, undirected and homogeneous networks. There exists works which exploit several network attributes to predict links. Few examples are: temporal link prediction~\cite{sett2018temporal,qin2023temporal} which exploits temporal dynamics of link maintenance in dynamic networks like in social networks; link prediction in bipartite networks~\cite{kunegis2010link,ozer2024link} which applies on bipartite graphs like user-product networks or term-document networks; link prediction in heterogeneous networks~\cite{sett2018temporal,li2018link,wang2023multi} which exploits the heterogeneity of nodes and links in networks like bibliographic networks; link prediction in directed networks~\cite{schall2014link,sett2018exploiting} which exploits the direction information in directed networks like email networks, etc. These specialized methods are not necessarily disjoint, and often use information covering multiple groups. In this paper, we consider the start-of-the-art LP methods only in our analysis, and reserve the specialized methods for future study.

Studies like~\cite{lichtnwalter2012link,yang2015evaluating} investigated the issues associated with LP evaluation in considerable details. They stressed on an imbalanced sampling of test set, and advocated for AUPR and the PR curve to tackle class imbalance in evaluation. Yang et al.~\cite{yang2015evaluating} also proposed a test set-up for supervised future LP from deployment perspective, and analyzed the effect of variation of the test node pair collection time duration. He et al.~\cite{he2024link} investigated how LP accuracy in real-world networks gets affected under non-uniform missing-edge patterns. A few studies like~\cite{junuthula2016evaluating,poursafaei2022towards} investigated the evaluation problem for temporal link prediction. These studies consider the timestamps of test node pairs as an important aspect for evaluation. Kumar et al.~\cite{kumar2020link} presents a nice survey on LP methods and evaluate them using standard metrics like AUROC, AUPR and Precision@K. Mostly, the existing studies on link prediction evaluation deal with the class imbalance issue~\cite{lichtnwalter2012link,yang2015evaluating,kumar2020link,lichtenwalter2010new,masrour2020bursting,nasiri2023robust}, but do not address the early retrieval performance. To the best of our knowledge, there exists no study which considers the multifaceted aspects of LP evaluation in such a detailed manner like us.
\section{Link Prediction Methods}
\label{LP-methods}
\subsection{Similarity Based Methods:}
Similarity-based LP methods calculate a score given a pair of nodes, which represents the likelihood of a link formation between them. Based on the node pair's geodesic distance in the network, similarity based methods are classified into two: \textit{local} and \textit{global}. Local methods are applied on the node pairs which are two-hop apart, but global methods can be applied on any node pair irrespective of their distance in the network. As the machine learning based methods also are global in nature, for ease of understanding, we refer local similarity based methods as \textit{local-sim} and global similarity based methods as \textit{global-sim} now onward. We summarize a few popular local and global similarity based methods which we use in this paper as follows.
\subsubsection{Local methods}\hfill \break
\textbf{Common Neighbors (\textit{CN})~\cite{liben2007link}:} The Common Neighbors method (\textit{CN}) measures the number of common neighbors or two-hop paths between node pairs. The $CN$ score between nodes $x$ and $y$ can be expressed as:
\begin{center}
\begin{equation*}
CN(x, y) = \mid\Gamma(x)\cap\Gamma(y)\mid,
\end{equation*}
\end{center}
where $\Gamma(x)$ represents the set of neighbors of $x$. \hfill \break
\textbf{Jaccord's Coefficient (\textit{JC})~\cite{dillon1983introduction,liben2007link}:} Jaccord's Coefficient (\textit{JC}) extends the $CN$ method by penalizing it for non-shared neighbors between the nodes. The $JC$ score between nodes $x$ and $y$ can be expressed as:
\begin{center}
\begin{equation*}
 JC(x, y) = \frac{\mid\Gamma(x)\cap\Gamma(y)\mid}  {\mid\Gamma(x)\cup\Gamma(y)\mid}
\end{equation*}
\end{center} \hfill \break
\textbf{Adamic Adar (\textit{AA})~\cite{liben2007link,admic2001friends}:} Adamic Adar (\textit{AA}) extends the $CN$ method by penalizing each common neighbor by its degree logarithmically. The $AA$ score between nodes $x$ and $y$ can be expressed as:
\begin{center}
\begin{equation*}
    AA\left(x, y\right) = \sum_{z \in \Gamma(x)\cap\Gamma(y)} \frac{1}{\log |\Gamma(z)|}
\end{equation*}
\end{center}\hfill \break
\textbf{Resource Allocation (\textit{RA})~\cite{zhou2009predicting}:} Unlike $AA$, Resource Allocation (\textit{RA}) penalizes each common neighbor by its degree without logarithmic scaling. The $RA$ score between nodes $x$ and $y$ can be expressed as:
\begin{center}
\begin{equation*}
    RA\left(x, y\right) = \sum_{z \in \Gamma(x)\cap\Gamma(y)} \frac{1}{|\Gamma(z)|}
\end{equation*}
\end{center}
\subsubsection{Global methods}\hfill \break
\textbf{Preferential Attachment (\textit{PA})~\cite{barabasi2002evolution,newman2001clustering,liben2007link}:}
Preferential Attachment method (\textit{PA}) relies on the principle that nodes with higher degrees are more likely to acquire new connections. The $PA$ score between nodes $x$ and $y$ can be expressed as:
\begin{center}
\begin{equation*}
  PA(x, y) = |\Gamma(x)| \times |\Gamma(x)|
\end{equation*}
\end{center}
\textbf{Katz Similarity (\textit{Katz})~\cite{katz1953new,liben2007link}:}
Katz similarity index (\textit{Katz}) enumerates all the possible paths of different lengths between the node pairs, and takes sum over this collection, exponentially damping by path length. The \textit{Katz} score between nodes $x$ and $y$ can be expressed as:
\begin{center}
\begin{equation*}
Katz(x, y) = \displaystyle \sum^\infty_{l=1} \beta^l\cdot\mid paths^{\left<l\right>}_{x,y}\mid,
\end{equation*}
\end{center}
where $\beta$ acts as decay factor to give exponentially higher weight to longer paths, and $\mid paths^{\left<l\right>}_{x,y}\mid$ is the number of different paths of length $l$ connecting the node pair.\hfill \break
\textbf{Hitting Time (\textit{HT}) and Normalized \textit{HT} (\textit{Norm-HT})~\cite{liben2007link}:}
Hitting Time (\textit{HT}) leverages random walks on a network to quantify node similarity. The hitting time between nodes $x$ and $y$ is the expected number of steps it takes for a random walker starting at node $x$ to reach node $y$ for the first time. It quantifies how easily information or influence can spread between the nodes. It indicates easier information flow or shorter travel times between the nodes in the network, where smaller \textit{HT} indicates better link formation likelihood. The scoring function can be expressed as:
\begin{center}
\begin{equation*}
HT(x, y)=-\displaystyle \sum^\infty_{t=1} t.P(T_{xy}=t)
\end{equation*}
\end{center}
Here, $T_{xy}$ is the random variable denoting the time it takes for a random walker to reach node $y$ from $x$, and $P(T_{xy}=t)$ is the probability of this happening in $t$ steps. \hfill \break
$HT(x, y)$ is quite small when the node $y$ has a large stationary probability. To mitigate this issue, the score is multiplied with $y$'s stationary probability. We call this measure as normalized heating time (\textit{Norm-HT}).\break
\textbf{Commute Time (\textit{CT}) and Normalized \textit{CT} (\textit{Norm-CT})~\cite{liben2007link}:}
Commute Time (\textit{CT}) signifies the expected time for a random walker to travel from node $x$ to $y$ and back to $x$. It encapsulates the symmetric nature of node connectivity. The \textit{CT} score between nodes $x$ and $y$ can be expressed as:
\begin{center}
\begin{equation*}
    CT(x,y)=HT(x, y)+HT(y, x)
\end{equation*}
\end{center}
Like HT, we consider normalized commute time (\textit{Norm-CT}) along with its un-normalized version.
\subsection{Machine Learning Based Methods}
We use three popular node embedding methods in this category, namely, \textit{Deepwalk}~\cite{perozzi2014deepwalk}, \textit{Node2vec}~\cite{grover2016node2vec}, and \textit{Graphsage}~\cite{hamilton2017inductive}, towards LP. All of these methods learn a function $f : V \rightarrow \mathbb{R}^d$, given a network graph $G(V,E)$, where $V$ and $E$ are the set of vertices and edges respectively, which maps each node to a $d$ dimensional latent space where $d<<|V|$. We learn edge features following the method presented in Grover et al.~\cite{grover2016node2vec} to get a link vector given a node pair, and apply \textit{logistic regression} and \textit{random forest} supervised learning technique to predict links. We refer this group of LP methods as \textit{learning}. Below, we brief the aforementioned node embedding methods and the edge feature learning methods.
\subsubsection{Deepwalk~\cite{perozzi2014deepwalk}:} Deepwalk adapts \textit{Skip-gram}~\cite{mikolov2013distributed} method of natural language processing to generate node embeddings. It solves the following optimization problem:
\begin{center}
\begin{equation*}
    \underset{f}{\text{minimize}} - \sum_{v_i \in V} \log {Pr(\{v_{i-w} , \ldots , v_{i-1}, v_{i+1}, \ldots , v_{i+w} \} | f(v_i))},
\end{equation*}
\end{center}
where $\{v_{i-w} , \ldots , v_{i-1}, v_{i+1}, \ldots , v_{i+w} \}$ are the context vertices of the vertex $v_i$ within a window of size $w$. To get the context vertices, Deepwalk simulates multiple truncated random walks started at every vertex, and extracts the context vertices from the walks. It approximates the optimization by solving it with a conditional independence assumption on the vertices in the context window, given the embedding of a vertex. On this assumption, it models the conditional probability of each node-context pair with a softmax unit. The optimization equation can be given by:
\begin{center}
\begin{equation*}
    \underset{f}{\text{maximize}} \sum_{v_i \in V} \big[ - \log Z_{v_i} + \sum_{n_i \in S^i} f(n_i) \cdot f(v_i) \big],
\end{equation*}
\end{center}
where $Z_{v_i}=\sum_{u \in V} \text{exp}(f(u) \cdot f(v_i))$, and $S^i$ are the set of nodes inside the context window of $v_i$. To avoid the explosion of labels, precisely $|V|$ numbers, Deepwalk uses Hierarchical Softmax~\cite{morin2005hierarchical,mnih2008scalable} with stochastic gradient descent (SGD) to approximate the optimization.
\subsubsection{Node$2$vec~\cite{grover2016node2vec}:} Like Deepwalk, Node$2$vec adapts Skip-gram to generate the embedding, and solves a similar optimization problem. It differs from Deepwalk in two aspects: generating the context window, and approximating the optimization problem. It argues that embeddings generated using truncated random walk can not capture homophily and structural equivalence among nodes. In order to capture a more flexible contextual structure, it employs a biased random walk method that combines neighborhood exploration using breadth-first search (BFS) and depth-first search (DFS). The context sampled by BFS generates similar embeddings for structurally equivalent nodes. DFS can explore distant parts of the network, which leads the embeddings to preserve homophily. Like DeepWalk, with the conditional independence assumption over the context window, Node$2$vec models the conditional probability with a softmax units. Node$2$vec deals with the explosion of labels by negative sampling with SGD to approximate the optimization.
\subsubsection{GraphSAGE~\cite{hamilton2017inductive}:} GraphSAGE is a graph neural network approach for scalable and inductive learning on large graphs. GraphSAGE leverages node attributes (e.g., node2vec embeddings) to learn embedding functions that generalize to unseen nodes during the training phase. GraphSAGE does this by learning aggregator functions that can induce an embedding of a node by aggregating the attributes of its neighboring nodes, sampled from its direct connections. This aggregation process is executed $k$ times for all nodes in the network, which way it learns $k$ sets of weight matrices $\{W^k\}$. It uses four different aggregator techniques: \textit{mean, max-pooling, mean-pooling}, and \textit{LSTM}. GraphSAGE optimizes similar objective as Deepwalk and Node$2$vec, and approximates it with negative sampling, where given a node, its positive instances are sampled from nodes appearing in the chain of short random walks starting at the given node, and negatives are sampled from the degree distribution. It uses SGD as the optimization procedure.
\subsubsection{Learning edge features:} To learn edge features, we follow the procedure described in Grover et al. ~\cite{grover2016node2vec}. For two nodes $x$ and $y$, we employ a binary operator ``$\circ$'' for the feature vectors $f(x)$ and $f(y)$ in order to achieve an edge representation $g(x, y)$ such that $g : V \times V \rightarrow \mathbb{R}^{d'}$ where $d'$ is the representation size for the pair $(x, y)$. The operators designed for this purpose are designed to be adaptable, allowing them to be applied to any node pair, even if there is no existing edge. This flexibility makes representations more valuable, especially in LP settings when there are both true and false edges in the test set. We summarize the operators in Table~\ref{tab:edge_oper}.
\begin{center}
    \begin{table}
    \centering
    \begin{tabular}{|l|c|c|}
    \hline
    Operator&Symbol&Definition\\
    \hline
    Average&$\oplus$&$[f(x)\oplus f(y)]_i=\frac{f_i(x)+f_i(y)}{2}$\\
    Hadamard&$\odot$&$[f(x)\odot f(y)]_i=f_i(x)*f_i(y)$\\
    Weighted-L1&$\Vert \cdot \Vert_{\Bar{1}}$&$\Vert f(x)\cdot f(y)\Vert_{\Bar{1}i}=\vert f_i(x)-f_i(y)\vert$\\
    Weighted-L2&$\Vert \cdot \Vert_{\Bar{2}}$&$\Vert f(x)\cdot f(y)\Vert_{\Bar{2}i}=\vert f_i(x)-f_i(y)\vert^2$\\
    \hline
    \end{tabular}
     \caption{Binary operators for Learning Edge Features~\cite{grover2016node2vec}}
    \label{tab:edge_oper}
    \end{table}
\end{center}
\subsubsection{Naming convention of individual link prediction methods:} Now onward, we refer each of the similarity based methods with their abbreviations. There are multiple variants of GraphSAGE for different aggregation techniques. There are multiple variants of the LP methods using each of the embedding methods depending on the edge learning operator, and the supervised learning method used. We abbreviate individual LP methods generated from these embedding methods as: \textit{(embedding-method)\_(aggregation-technique-if-applicable)\_(edge-learning-operator)\_(supervised-method)}. To improve presentability, we further abbreviate Deepwalk, Node$2$vec and GraphSAGE as DW, N$2$V and GS respectively, the aggregator techniques mean, max-pooling, mean-pooling and LSTM as \textit{Mean, MeanPool, MaxPool} and \textit{LSTM} respectively, the edge learning operators Average, Hadamard, Weighted-L1 and Weighted-L2 as \textit{Avg, Hada, L1} and \textit{L2} respectively, and the supervised learning methods logistic regression and random forest as \textit{LR} and \textit{RF} respectively. For an example, \textit{GS\_Mean\_Avg\_RF} refers to the LP method which generates the node embeddings using Mean aggregator technique with GraphSAGE, and then learns edge features using the Average operator, and then trains a random forest model to predict links. In total, our experiments are conducted on 58 distinct LP methods, of which, 4 are \textit{local-sim}, 6 are \textit{global-sim}, and 48 are \textit{learning} based methods.
\section{Datasets}
\label{datasets}
We conducted our experiments on fifteen real network datasets of diverse nature, such as, social network, communication network, collaboration network, biological network, power grid, etc. A few of these networks like friendship networks, collaboration networks are undirected in nature, whereas networks like communication networks are directed. The social, communication and collaboration networks are inherently dynamic or temporal in nature. Among the networks in our collection which are dynamic, few had timestamps available with the interactions or the moment of link formation. In total, nine among all the networks had timestamps. For these networks we created ground truth labels two ways: (a) considering the timestamps which lets us evaluate future LP, and (b) removing a portion of the links in the networks which lets us evaluate the missing LP. So, we created two LP datasets for each of these networks. For the networks with no timestamp, we could only evaluate missing LP. In total, we conducted our experiments and analysis on $24$ datasets, among which 15 were used for missing LP and the rest were used for future LP. We summarize the datasets in Table \ref{tab:NW}. We briefly describe  the datasets as follows.

\subsection{Networks with timestamps:}
\begin{enumerate}
   \item \textbf{CollegeMsg~\cite{panzarasa2009patterns}:} This dataset represents a temporal network capturing private communications among college students on a campus at the University of California, Irvine. Each node in the network represents an individual student, while the private messages exchanged between them guides the link formation and maintenance. This dataset offers insights into communication dynamics and prevalent styles within the campus community. Information for this dataset was obtained over the course of 193 days at the University of California, Irvine. In this paper, we refer the dataset which prepares ground truth using timestamps as \texttt{clg-msg-t} and the the one which do not consider the timestamps as \texttt{clg-msg}.
   \item \textbf{Email-Eu~\cite{paranjape2017motifs}:} This dataset provides insight into the evolving connections and communication patterns within a European research organization over time. It primarily focuses on email correspondence exchanged between members of the institution. The network's nodes represent individual users or email addresses within the organization, and the links encode the email exchanges between them, with timestamps indicating the exact moments of interaction. This data enables researchers to analyze the dynamics of information flow and collaboration within the organization. Information for this dataset was gathered over a period of 803 days. In this paper, we refer the temporal version as \texttt{email-t} and the non-temporal one as \texttt{email}.
    \item \textbf{Mathoverflow~\cite{paranjape2017motifs}:} This dataset presents a temporal network capturing interactions among users on the Mathoverflow platform, a specialized community for mathematicians to exchange knowledge and insights. Each node in the network represents an individual user, while the links encode their interactions, such as comments or responses to questions and answers. The timestamps associated with the interactions indicate the exact moment of these interactions, providing valuable temporal insights into the dynamics of knowledge sharing and collaboration within the mathematical research community. This dataset is taken over a time span of 2350 days. In this paper, we refer the temporal version as \texttt{math-flow-t} and the non-temporal version as \texttt{math-flow}.
    \item \textbf{Ast-phys~\cite{leskovec2007graph}:} This dataset contains a collaboration network of authors who have contributed to the field of Astrophysics through scientific paper publications. Each node represents an individual author, and the links signify collaborations between authors. The timestamps indicate the time of collaborations between two authors. In this paper, we refer the temporal version as \texttt{ast-phys-t} and the non-temporal version as \texttt{ast-phys}.
    \item \textbf{LastFm~\cite{kumar2019predicting}:} This dataset represents a social network derived from \textit{Last.fm}, a music streaming and recommendation service. Nodes in the network represent Last.fm users, and the links symbolize their interactions and social relationships. The dataset includes timestamps indicating the precise moments of user connections and interactions. In this paper, we refer the temporal version as \texttt{lst-fm-t} and the non-tempral one as \texttt{lst-fm}.
    \item \textbf{IAcontact~\cite{nr}:} This dataset represents a contact network among attendees of the ACM Hypertext conference in the year 2009. The links encode the interactions among the attendees during a 2.5-day period, while the nodes represent the attendees. In this paper, we refer the temporal version as \texttt{iacontact-t} and the non-temporal one as \texttt{iacontact}.
    \item \textbf{FB-forum~\cite{nr}:} This dataset illustrates a forum-based online social network similar to Facebook. The network mostly records user actions and conversations in this forum. Individual nodes in this dataset represent users, and the links represent the forum messages these users have exchanged. The dataset spans a period of 164 days. In this paper, we refer the temporal version as \texttt{forum-t} and the non-temporal one as \texttt{forum}.
    \item \textbf{Topology~\cite{konect,konect:zhang05}:} This dataset encodes the complicated network of connections between autonomous systems on the Internet. Node refers to autonomous systems (AS), which are collections of connected IP routing domains run by independent network operators. In this paper, we refer the temporal version as \texttt{topo-t} and the non-temporal one as \texttt{topo}.
    \item \textbf{Mooc~\cite{kumar2019predicting}:} This dataset displays user activity on a well-known MOOC program. The actions are symbolized as a temporal and directed network. Users are represented as nodes in the graph together with course activities (targets), and the links are user actions on the targets. In this paper, we refer the temporal version as \texttt{act-mooc-t} and the non-temporal one as \texttt{act-mooc}.
\end{enumerate}

\subsection{Networks without timestamps:}
\begin{enumerate}
   \item \textbf{ArXiv~\cite{leskovec2007graph}:} ArXiv is a network dataset representing collaborations among scientists who publish preprints and papers on the ArXiv platform. Nodes in the network represent individual authors, while links represent co-authorship connections. We refer the dataset prepared from this network as \texttt{arxiv}.
   \item \textbf{Power-grid~\cite{nr}:} The Power-grid network dataset models the interconnected power grid infrastructure in the United States. It provides information on the transmission lines, nodes, and elements that make up the electrical system, enabling research into electricity distribution across different geographic areas. Nodes represent power plants, substations, transformers, and other infrastructure, while links depict the transmission lines connecting these elements. This dataset is valuable for studying the dynamics and efficiency of the U.S. electrical power grid. In this paper, we refer the dataset prepared from this network as \texttt{pow-grid}.
    \item \textbf{Tech-routers~\cite{nr}:} It is a tech routers network, which represents the connections between routers in a technological infrastructure like the internet or computer networks. Nodes in the network represent individual routers, and links symbolize direct physical or logical links between matching routers. We refer the dataset prepared from this network as \texttt{routers}.
    \item \textbf{Bio-yeast~\cite{nr}:} The Bio-yeast dataset is a biological network database focused on protein-protein interactions in the yeast species Saccharomyces cerevisiae. It represents proteins as nodes and shows connections between proteins based on physical binding, enzymatic processes, or co-expression patterns. We refer the dataset prepared from this network as \texttt{bio-yeast}.
    \item \textbf{Facebook~\cite{leskovec2012learning}:} It is a social network, built on the Facebook platform, focuses on a single user (``ego") and their connections with other users. Nodes represent Facebook users, and links represent friendships between them. We refer the dataset prepared from this network as \texttt{fb}.
    \item \textbf{BlogCatalog~\cite{nr}:} This dataset is a social relationship network. The network is made up of bloggers and their friends and other social connections. Nodes in this network are represented as bloggers and the links represent social relationship among the bloggers. In this paper, we refer the dataset prepared from this network as \texttt{blog-cat}.
\end{enumerate}

\begin{sidewaystable}
\caption{Dataset information}
\label{tab:NW}
\begin{tabular*}{\textwidth}{@{\extracolsep\fill}lcclclccc}
\hline
\textbf{Network} & \textbf{\#Nodes} & \textbf{\#Edges} & \textbf{Type} & \textbf{Time Span} & \textbf{Dataset} & \textbf{\#Positives} & \textbf{\makecell{\#two-hop\\ positives}} & \textbf{\makecell{\#three-hop\\ positives}} \\ \midrule
\multirow{2}{*} {CollegeMsg} & \multirow{2}{*}{1899} & \multirow{2}{*}{20296} & \multirow{2}{*}{Directed} & \multirow{2}{*}{194 days} & \texttt{clg-msg-t} & 2142 & 1068(49.80\%) & 983(45.89\%) \\
\cmidrule{6-9}
 &  &  &  &  & \texttt{clg-msg} & 510 & 335(65.68\%) & 174(34.11\%) \\ \midrule
\multirow{2}{*}{Email-Eu} & \multirow{2}{*}{986} & \multirow{2}{*}{332334} & \multirow{2}{*}{Directed} & \multirow{2}{*}{803 days} & \texttt{email-t} & 3977 & 3774(94.89\%) & 200(5.03\%) \\
\cmidrule{6-9}
 &  &  &  &  & \texttt{email} & 441 & 429(97.27\%) & 11(2.49\%) \\ \midrule
\multirow{2}{*}{Mathoverflow} & \multirow{2}{*}{24818} & \multirow{2}{*}{506550} & \multirow{2}{*}{Directed} & \multirow{2}{*}{2350 days} & \texttt{math-flow-t} & 24225 & 18918(78.09\%) & 4889(20.18\%) \\
\cmidrule{6-9}
 &  &  &  &  & \texttt{math-flow} & 11373 & 9981(87.76\%) & 1125(9.89\%) \\ \hline
\multirow{2}{*}{Ast-Phys} & \multirow{2}{*}{9756} & \multirow{2}{*}{86697} & \multirow{2}{*}{Undirected} & \multirow{2}{*}{3316 days} & \texttt{ast-phys-t} & 32186 & 27035(83.99\%) & 5151(16\%) \\
\cmidrule{6-9}
 &  &  &  &  & \texttt{ast-phys} & 8898 & 8862(99.59\%) & 36(0.40\%) \\ \midrule
 \multirow{2}{*}{LastFm} & \multirow{2}{*}{7624} & \multirow{2}{*}{27806} & \multirow{2}{*}{Undirected} & \multirow{2}{*}{1345 days} & \texttt{lst-fm-t} & 29867 & 29862(99.38\%) & 185(0.61\%)  \\
\cmidrule{6-9}
 &  &  &  &  & \texttt{lst-fm} & 3755 & 3748(99.81\%) & \--- \\ \midrule
 \multirow{2}{*}{IAcontact} & \multirow{2}{*}{113} & \multirow{2}{*}{2498} & \multirow{2}{*}{Directed} & \multirow{2}{*}{1 day} & \texttt{iacontact-t} & 533 & 533(100\%) & \--- \\
\cmidrule{6-9}
 &  &  &  &  & \texttt{iacontact} & 92 & 92(100\%) & \--- \\ \midrule
\multirow{2}{*}{FB-forum} & \multirow{2}{*}{899} & \multirow{2}{*}{7089} & \multirow{2}{*}{Directed} & \multirow{2}{*}{164 days} & \texttt{forum-t} & 1638 & 717(43.58\%) & 862(52.81\%) \\
\cmidrule{6-9}
 &  &  &  &  & \texttt{forum} & 277 & 163(58.84\%) & 109(39.35\%) \\ \midrule
 \multirow{2}{*}{Topology} & \multirow{2}{*}{34761} & \multirow{2}{*}{171403} & \multirow{2}{*}{Undirected} & 23 days & \texttt{topo-t} & 21100 & 17331(82.13\%) & 3159(14.97\%) \\
\cmidrule{6-9}
 &  &  &  & \- & \texttt{topo} & 4682 & 3578(76.42\%) & 980(20.93\%) \\ \midrule

\multirow{2}{*}{Mooc} & \multirow{2}{*}{7047} & \multirow{2}{*}{411749} & \multirow{2}{*}{Directed} & \multirow{2}{*}{29 days} & \texttt{act-mooc-t} & 38448 & 38381(99.82\%) & 67(0.01\%) \\
\cmidrule{6-9}
 &  &  &  &  & \texttt{act-mooc} & 8594 & 8590(99.95\%) & \--- \\ \midrule
ArXiv & 23133 & 93133 & Undirected & \- & \texttt{arxiv} & 886 & 867(97.85\%) & 6(0.14\%) \\ \midrule
Power-grid & 4941 & 6594 & Undirected & \- & \texttt{pow-grid} & 183 & 117(63.93\%) & 65(35.51\%) \\ \midrule
Tech-routers & 2113 & 6632 & Undirected & \- & \texttt{routers} & 590 & 72(12.20\%) & 45(7.62\%) \\ \midrule
Bio-yeast & 6008 & 156945 & Undirected & \- & \texttt{bio-yeast} & 1572 & 1472(93.63\%) & 100(6.36\%) \\ \midrule
Facebook & 4039 & 88234 & Undirected & \- & \texttt{fb} & 8813 & 8796(99.80\%) & 17(0.19\%) \\ \midrule
BlogCatalog & 10312 & 333983 & Undirected & \- & \texttt{blog-cat} & 33350 & 33112(99.28\%) & 238(0.71\%) \\
\bottomrule
\end{tabular*}
\end{sidewaystable}

\section{Experimental Setup}
\label{exp-setup}
\subsection{Graph and ground truth generation}
\label{exp-setup-gt}
In this section, we detail the procedure of how we prepare the graph and the ground truth positive and negative node pairs from the datasets presented in Section~\ref{datasets}, for future and missing LP tasks. Unlike traditional approaches, we carefully segregate the test examples based on the distance of the test node pairs in the graph, and maintain the distance bias inherent with the positive examples' distribution, while sampling the negative examples. Moreover, it takes care of the evaluation in a class imbalance environment and its interactions with the factors stated above.

The future LP datasets were prepared for the networks where the interaction or link formation timestamps were available. The interactions can be directed, and there could be multiple interactions between a node pair. As we aim to prepare a simple and undirected graph and the true future links from a network, we ignored the directions of the interactions and the duplicate interactions. The detailed procedure for the dataset preparation is described as follows. We first sort the interactions or links (for the cases where only the link formation timestamps were available) by their timestamp values, and then choose $75 \%$ of the earlier timestamped interactions or links for constructing the graph, and the remaining were used for generating positive examples. From the interactions which are in that $75 \%$, we constructed an unweighted and undirected graph by first ignoring the directions and then removing duplicate interactions and links, if any. If the graph is having multiple connected components, we choose the largest connected component as the final graph for our experiment. From the remaining $25 \%$ interactions, we ignored the directions and removed the duplicates to generate the potential positive examples. From this potential set, we first ignored the node pairs which involve at least one node which is not included in the graph, and then removed the links which are present in the graph as an edge to get the final set of positive examples.

The missing LP was applied to all the networks presented in Section~\ref{datasets}. We preprocessed the networks having timestamp information by ignoring the timestamps and their directions, and then removing the duplicates. Similarly, we ignored the directions and removed the duplicates, if applicable, for the networks which did not have timestamp information. After doing this, we are left with a set of node pairs which represents links. Then we randomly remove $10 \%$ of the links from the set. We construct the graph with the remaining links and choose the largest connected component as the final graph. We consider the removed $10 \%$ links as the potential positive examples, and generate the positive examples following the same procedure as the future LP datasets.

For each dataset, we divided the positive examples in three parts based on the shortest path distance of the end node pair in the graph: having the shortest path distance (a) $2$, (b) $3$, and (c) more than $3$. As most of the datasets have no or a very few end node pairs having the shortest path distance $>3$, we do not consider this group in our experiment. To evaluate the effect of class imbalance on the LP methods, we generated negative examples $10$ times as many as the number of positive examples, stratified by specific shortest path distance categories maintaining the ratio of the representatives of the categories same as the positive set, and conducted experiments on multiple skew levels. For each category, we repeatedly choose two random nodes having the specified shortest path distance in the graph and if they are not included in the set of positive examples, we include the pair in the negative set. This sampling strategy empower us to evaluate the LP methods in groups of test examples by their shortest path distance in class imbalanced setup. The details of the positive example count for each dataset grouped by the distance categories are provided in Table~\ref{tab:NW}.
\subsection{Hyper-parameter settings for the LP methods}
The Katz similarity method was implemented with $\beta = 0.05$, which is the typical value used in the literature~\cite{liben2007link}. Hyper-parameters of the node embedding models: Deepwalk~\cite{perozzi2014deepwalk}, Node$2$vec~\cite{grover2016node2vec}, GraphSAGE~\cite{hamilton2017inductive}, were set following the recommendations given in the actual paper. We performed five-fold cross validation, stratified by the skew level of the class distribution, to apply logistic regression and random forest on the link features towards predicting links.
\subsection{Evaluation measures}
The area under the receiver operating curve (AUROC)~\cite{hanley1982meaning} has historically been the primary performance metric used to assess the performance of LP methods~\cite{kumar2020link,martinez2016survey}. The receiver operating curve (ROC) plots the true positive rate (TPR) against the false positive rate (FPR) at every threshold value. AUROC gives the probability of a positive example assigned a higher score than that of a negative example. However, AUROC is not a suitable performance measure in a class imbalance scenario like LP problem, which is inherently skewed towards the negative class. AUROC favors accurate classification of positive examples, at the cost of significantly misclassifying negative examples. Hence, AUROC can overestimate the performance of LP. An alternative to AUROC is area under the precision-recall curve (AUPR)~\cite{saito2015precision,lichtenwalter2010new}. Precision-recall (PR) curve plots precision at every threshold value vs. its recall, i.e., TPR. As precision is sensitive to class imbalance unlike FPR, it can provide better evaluation of LP~\cite{saito2015precision}. Despite AUPR's suitability, it is less popular in LP literature, may be due to the variability of PR curves in different skew levels, which makes it difficult to compare the LP methods when multiple skewness is considered. To deal with class imbalance, another skew sensitive evaluation measure, $Precision@k$ is sometimes used to evaluate LP as a single threshold measure, where the positive example count is considered as a typical value of $k$~\cite{kerrache2020scalable,kumar2020link}.

In this work, we consider AUROC, AUPR and $Precision@k$ as single point summary metrics for evaluating LP methods. For the skew sensitive metrices: AUPR and $Precision@k$, we consider both balanced as well as imbalanced (skew level $1:10$) test set. We choose the $k$ value to be the positive example count, and name it as $Pr@P$. To understand the behavior of the LP methods in early retrieval stage, and the effect of class imbalance on it, we employ $Precision@k$ setting $k$ value to be the half of the positive example count, which we denote as $Pr@P/2$. We also use the PR curve (PRC) along with it to understand the behavior covering all threshold values.

\section{Results and Discussion}
\label{Res_Dis}
We applied the 58 LP methods presented in Section~\ref{LP-methods} on the 24 datasets defined in Section~\ref{datasets} in a controlled experimental setup proposed in Section~\ref{exp-setup}. We evaluated their performance based on the four evaluation metrics: AUROC, AUPR, $Pr@P$ and $Pr@P/2$ on the test sets. As described in Section~\ref{exp-setup-gt}, we had different test sets depending on the shortest path distance between test node pairs, and each of these had two versions: balanced and imbalanced. For the skew sensitive metrics: AUPR, $Pr@P$ and $Pr@P/2$, we consider evaluating both for balanced and imbalanced test set. We performed a series of statistical tests to test an array of carefully designed hypotheses in this controlled setup to interrogate how the LP methods perform in different situations depending on the shortest path distance between test node pairs, prediction type, network type, and class imbalance. We also investigate how the class imbalance affect the LP methods in early retrieval phase using proper statistical tests and PRC plots. We present the detailed results along with the insights acquired from them as follows.
\subsection{Future LP Vs. Missing LP}
In this section, we investigate whether the prediction performance of the LP methods differ on a particular network for the two cases: future LP and missing LP. Given a network, we apply paired t-test to compare the absolute values of the prediction performances of the prediction methods given a particular evaluation metric for the two cases. To find the level of the ordinal association between the two cases, we apply Kendall?s Tau Rank Correlation Coefficient ($\tau$). We consider AUROC, and both balanced and imbalanced versions of AUPR and $Pr@P$ as the evaluation measure. We perform the experiment on the nine networks which had timestamps in distance controlled test set. Table \ref{tab:FutvsMis-Pttest} and Table \ref{tab:FutvsMis-Kend} summarize the results of the two tests respectively.

Almost all paired t-test p-values are significant across distances and evaluation measures, and for almost all networks, the LP methods perform better for future LP than missing LP. The only exception is the two-hop case of LastFm network, where either the missing LP wins or there is no significant difference. When the test node pairs are generated randomly in a longitudinal network, the temporal relationship among the links are lost. The outperformance of future LP might be due to the adverse affect of this information loss in their missing LP counterparts. This also indicates that when an application demands future LP, but the ground truth links are generated in traditional way, i.e., removing a portion of links from the network randomly, the prediction performance is significantly underestimated, and vice versa. Table \ref{tab:FutvsMis-Kend} shows that except for the Topology and Mooc network, all rank correlation values are extremely low, in few cases even negative. This indicates that, for a future LP, if the ground truth links are generated in traditional way, the prediction performance rankings of various LP methods vary significantly than if the ground truth links were generated with true future links. Our results indicate the necessity to evaluate a LP method in both setups and report its performance in both types of prediction tasks, in order to understand its effectiveness.
\begin{sidewaystable}
\caption{Paired t-test results for future LP vs missing LP}
\begin{tabular*}{\textwidth}{@{\extracolsep\fill}lllclclclclcl}
\toprule
\multirow{3}{*}{} & \multirow{3}{*}{\textbf{NETWORK}} & \multicolumn{2}{c}{\multirow{3}{*}{\textbf{AUROC}}} & \multicolumn{4}{c}{\textbf{AUPR}} & \multicolumn{4}{c}{\textbf{Pr@P}} \\
\cmidrule(r){5-8} \cmidrule(r){9-12}

 &  & \multicolumn{2}{c}{} & \multicolumn{2}{c}{\textbf{balanced}} & \multicolumn{2}{c}{\textbf{imbalanced}} & \multicolumn{2}{c}{\textbf{balanced}} & \multicolumn{2}{c}{\textbf{imbalanced}} \\
\cmidrule(r){3-4} \cmidrule(r){5-6} \cmidrule(r){7-8} \cmidrule(r){9-10} \cmidrule(r){11-12}
 &  & \textbf{P-Value} & \textbf{Winner} & \textbf{P-Value} & \textbf{Winner} & \textbf{P-Value} & \textbf{Winner} & \textbf{P-Value} & \textbf{Winner} & \textbf{P-Value} & \textbf{Winner} \\
\midrule
\multirow{9}{*}{\begin{sideways}two-hop\end{sideways}} & CollegeMsg & $<$.001*** & future & $<$.001*** & future & $<$.001*** & future & $<$.001*** & future & $<$.001*** & future \\
& Email-Eu & $<$.001*** & future & $<$.001*** & future & $<$.001*** & future & $<$.001*** & future & $<$.001*** & future \\
& Mathoverflow & $<$.001*** & future & $<$.001*** & future & $<$.001*** & future & $<$.001*** & future & $<$.001*** & future \\
& Ast-phys & $<$.001*** & future & $<$.001*** & future & 0.003** & future & $<$.001*** & future & 0.003** & future \\
& LastFm & $<$.001*** & missing & 0.105 & --- & 0.091 & --- & $<$.001*** & missing & $<$.001*** & missing \\
& IAcontact & $<$.001*** & future & $<$.001*** & future & $<$.001*** & future & $<$.001*** & future & $<$.001*** & future \\
& FB-forum & $<$.001*** & future & $<$.001*** & future & 0.052 & --- & $<$.001*** & future & $<$.001*** & future \\
& Topology & $<$.001*** & future & $<$.001*** & future & $<$.001*** & future & $<$.001*** & future & $<$.001*** & future \\
& Mooc & 0.011* & future & 0.018* & future & 0.004** & future & 0.006** & future & 0.004** & future \\
\midrule
\multirow{8}{*}{\begin{sideways}three-hop\end{sideways}} & CollegeMsg & $<$.001*** & future & $<$.001*** & future & 0.943 & --- & $<$.001*** & future & 0.346 & --- \\
& Email-Eu & $<$.001*** & future & $<$.001*** & future & $<$.001*** & future & $<$.001*** & future & $<$.001*** & future \\
& Mathoverflow & $<$.001*** & future & $<$.001*** & future & $<$.001*** & future & $<$.001*** & missing & $<$.001*** & future \\
& Ast-phys & $<$.001*** & future & $<$.001*** & future & $<$.001*** & future & 0.298 & --- & $<$.001*** & future \\
& LastFm & --- & --- & --- & --- & --- & --- & --- & --- & --- & --- \\
& IAcontact & --- & --- & --- & --- & --- & --- & --- & --- & --- & --- \\
& FB-forum & $<$.001*** & future & $<$.001*** & future & $<$.001*** & future & $<$.001*** & future & $<$.001*** & future \\
& Topology & $<$.001*** & future & $<$.001*** & future & $<$.001*** & future & 0.003** & future & $<$.001*** & future \\
& Mooc & --- & --- & --- & --- & --- & --- & --- & --- & --- & --- \\
\bottomrule
\end{tabular*}
\label{tab:FutvsMis-Pttest}
\end{sidewaystable}

\begin{center}
\begin{table*}[!h]%
\caption{Kendall's $\tau$ correlation results for future LP vs missing LP.\label{tab:FutvsMis-Kend}}
\begin{tabular*}{\textwidth}{@{\extracolsep\fill}llccccc@{}}
\toprule
\multirow{2}{*}{} & \multirow{2}{*}{\textbf{NETWORK}} & \multirow{2}{*}{\textbf{AUROC}} & \multicolumn{2}{c}{\textbf{AUPR}} & \multicolumn{2}{c}{\textbf{Pr@P}} \\ \cmidrule(r){4-5} \cmidrule(r){6-7}
    &    & & \textbf{balanced} & \textbf{imbalanced} & \textbf{balanced} & \textbf{imbalanced} \\
\midrule
\multirow{9}{*}{\begin{sideways}two-hop\end{sideways}} &CollegeMsg & 0.184 & 0.184 & 0.219 & 0.156 & 0.203 \\
& Email-Eu & 0.272 & 0.319 & 0.472 & 0.294 & 0.526 \\
& Mathoverflow & 0.339 & 0.386 & 0.414 & 0.326 & 0.488 \\
& Ast-phys & 0.487 & 0.433 & 0.377 & 0.525 & 0.393 \\
&  lastFm & 0.477 & 0.465 & 0.381 & 0.462 & 0.335 \\
&  IAcontact & 0.027 & 0.080 & 0.038 & 0.039 & 0.075 \\
&  FB-forum & 0.321 & 0.347 & 0.438 & 0.311 & 0.271 \\
&  Topology & 0.785 & 0.783 & 0.771 & 0.723 & 0.742 \\
&  Mooc & 0.744 & 0.743 & 0.698 & 0.678 & 0.741 \\
\midrule
\multirow{8}{*}{\begin{sideways}three-hop\end{sideways}}
&CollegeMsg & 0.009 & 0.118 & 0.158 & 0.022 & 0.230 \\
& Email-Eu & -0.030 & 0.085 & 0.167 & -0.144 & 0.181 \\
&  Mathoverflow & -0.020 & -0.071 & 0.018 & 0.139 & 0.060 \\
&  Ast-phys & 0.256 & 0.464 & 0.578 & 0.422 & 0.628 \\
&  lastFm & --- & --- & --- & --- & --- \\
&  IAcontact & --- & --- & --- & --- & --- \\
&  FB-forum & 0.189 & 0.208 & 0.283 & 0.203 & 0.334 \\
&  Topology & 0.749 & 0.747 & 0.738 & 0.649 & 0.773 \\
&  Mooc & --- & --- & --- & --- & --- \\
\bottomrule
\end{tabular*}

\end{table*}
\end{center}

\subsection{Distance between the end nodes in a test node pair}
Here we aim to understand whether prediction performance of various LP methods vary when the distance between the two nodes in test node pairs differ. For each dataset, we performed paired t-test to determine if the prediction performance differ in terms of their absolute values for the two cases: two-hop away test node pairs and three-hop away test node pairs. Also, we measure the Kendall?s $\tau$ values for the the same to estimate the ordinal association between the two cases. We consider AUROC, AUPR (balanced), AUPR (imbalanced), $Pr@P$ (balanced) and $Pr@P$ (imbalanced) as the evaluation measures. We performed the experiment across all datasets and prediction types. We summarize the results of the two tests in Table \ref{tab:hop-Pttest} and Table~\ref{tab:hop-Kend} respectively.

Table~\ref{tab:hop-Pttest} shows that the absolute performance of the prediction methods differs significantly in most of the cases across all datasets, prediction types and evaluation metrics. Overall, for a given dataset, the winner between two-hop and three-hop is consistent across the evaluation measures. As far as the missing LP is concerned, there is no clear winner between the two cases. However, for the future LP, three-hop seems to do better than its two-hop counterpart, where, three-hop wins in \texttt{clg-msg-t}, \texttt{math-flow-t}, \texttt{ast-phys-t} and \texttt{lst-fm-t}, but there is no clear winner for the rest three datasets. When we compare the performance for the networks (wherever applicable) based on prediction type, we find some inconsistencies, which is most prominent in the Ast-phys network, where its missing LP version is dominated by two-hop, but three-hop does better in its future LP counterpart. These results indicate that the LP methods produce different level of accuracy for predicting links based on the distance of the end nodes in the network. Moreover, this behavior differs over the prediction types. From Table~\ref{tab:hop-Kend} it is also evident that the rankings of the prediction methods for the two cases vary significantly. Overall, the obtained results support the need to evaluate LP methods separately for the two-hop and three-hop case to understand its behavior in distance controlled manner, and our experimental setup provides an opportunity to measure it. Moreover, as the local LP methods are applicable only on in the two-hop case, but the global methods are applicable in both the cases, our experimental setup provides a robust environment to compare a local with a global method.

\begin{sidewaystable}
\caption{Paired t-test results for two-hop vs three-hop}

\begin{tabular*}{\textwidth}{@{\extracolsep\fill}llclclclclcl}
\toprule
\multirow{3}{*}{} & \multirow{3}{*}{\textbf{DATASET}} & \multicolumn{2}{c}{\multirow{3}{*}{\textbf{AUROC}}} & \multicolumn{4}{c}{\textbf{AUPR}} & \multicolumn{4}{c}{\textbf{Pr@P}} \\
\cmidrule(r){5-8} \cmidrule(r){9-12}

 &  & \multicolumn{2}{c}{} & \multicolumn{2}{c}{\textbf{balanced}} & \multicolumn{2}{c}{\textbf{imbalanced}} & \multicolumn{2}{c}{\textbf{balanced}} & \multicolumn{2}{c}{\textbf{imbalanced}} \\
\cmidrule(r){3-4} \cmidrule(r){5-6} \cmidrule(r){7-8} \cmidrule(r){9-10} \cmidrule(r){11-12}
 &  & \textbf{P-Value} & \textbf{Winner} & \textbf{P-Value} & \textbf{Winner} & \textbf{P-Value} & \textbf{Winner} & \textbf{P-Value} & \textbf{Winner} & \textbf{P-Value} & \textbf{Winner} \\
\midrule

\multirow{13}{*}{\begin{sideways}missing LP\end{sideways}}
& \texttt{clg-msg} & $<$.001*** & 3-hop & $<$.001*** & 3-hop & $<$.001*** & 3-hop & $<$.001*** & 3-hop & $<$.001*** & 3-hop \\
& \texttt{email} & $<$.001*** & 2-hop & $<$.001*** & 2-hop & $<$.001*** & 2-hop & $<$.001*** & 2-hop & $<$.001*** & 2-hop \\
& \texttt{math-flow} & 0.226 & --- & 0.223 & --- & 0.492 & --- & $<$.001*** & 3-hop & 0.170 & --- \\
& \texttt{ast-phys} & $<$.001*** & 2-hop & $<$.001*** & 2-hop & $<$.001*** & 2-hop & $<$.001*** & 3-hop & $<$.001*** & 2-hop \\
& \texttt{lst-fm} & --- & --- & --- & --- & --- & --- & --- & --- & --- & --- \\
& \texttt{forum} & 0.007** & 2-hop & 0.134 & --- & $<$.001*** & 2-hop & 0.012* & 2-hop & 0.066 & --- \\
& \texttt{topo} & $<$.001*** & 3-hop & $<$.001*** & 3-hop & $<$.001*** & 3-hop & $<$.001*** & 3-hop & $<$.001*** & 3-hop \\
& \texttt{act-mooc} & --- & --- & --- & --- & --- & --- & --- & --- & --- & --- \\
& \texttt{pow-grid} & 0.512 & --- & $<$.001*** & 2-hop & $<$.001*** & 2-hop & $<$.001*** & 2-hop & $<$.001*** & 2-hop \\
& \texttt{routers} & 0.923 & --- & $<$.001*** & 2-hop & 0.229 & --- & $<$.001*** & 2-hop & 0.029* & 3-hop \\
& \texttt{bio-yeast} & $<$.001*** & 2-hop & $<$.001*** & 2-hop & $<$.001*** & 2-hop & $<$.001*** & 3-hop & $<$.001*** & 2-hop \\
& \texttt{fb} & $<$.001*** & 3-hop & $<$.001*** & 2-hop & 0.101 & --- & $<$.001*** & 3-hop & 0.682 & --- \\
& \texttt{blog-cat} & $<$.001*** & 2-hop & $<$.001*** & 2-hop & $<$.001*** & 2-hop & $<$.001*** & 3-hop & $<$.001*** & 2-hop\\
\midrule
\multirow{8}{*}{\begin{sideways}future LP\end{sideways}}
& \texttt{clg-msg-t} & $<$.001*** & 3-hop & $<$.008** & 3-hop & $<$.007** & 3-hop & $<$.005** & 3-hop & $<$.004** & 3-hop \\
& \texttt{email-t} & $<$.001*** & 2-hop & 0.463 & --- & 0.448 & --- & $<$.001*** & 2-hop & $<$.049* & 3-hop \\
& \texttt{math-flow-t} & $<$.001*** & 3-hop & $<$.001*** & 3-hop & $<$.001*** & 3-hop & $<$.001*** & 3-hop & $<$.001*** & 3-hop \\
& \texttt{ast-phys-t} & 0.081 & --- & $<$.001*** & 3-hop & $<$.001*** & 3-hop & $<$.001*** & 3-hop & $<$.001*** & 3-hop \\
& \texttt{lst-fm-t} & $<$.001*** & 3-hop & $<$.001*** & 3-hop & $<$.001*** & 3-hop & $<$.001*** & 3-hop & $<$.001*** & 3-hop \\
& \texttt{forum-t} & 0.827 & --- & 0.255 & --- & 0.531 & --- & 0.563 & --- & 0.212 & --- \\
& \texttt{topo-t} & 0.334 & --- & 0.585 & --- & 0.004** & 3-hop & 0.017* & 2-hop & 0.013* & 3-hop \\
& \texttt{act-mooc-t} & 0.265& --- & $<$.004** & 2-hop & 0.824 & --- & $<$.001*** & 3-hop & 0.372 & --- \\
\bottomrule
\end{tabular*}

\label{tab:hop-Pttest}
\end{sidewaystable}

\begin{table}[ht]
\caption{Kendall's $\tau$ correlation results for two-hop vs three-hop}

\begin{tabular*}{\textwidth}{@{\extracolsep\fill}llccccc}
\toprule
\multirow{2}{*}{} & \multirow{2}{*}{DATASET} & \multirow{2}{*}{AUROC} & \multicolumn{2}{c}{AUPR} & \multicolumn{2}{c}{Pr@P} \\ \cmidrule(r){4-5} \cmidrule(r){6-7}
    &    & & balanced & imbalanced & balanced & imbalanced \\
\midrule
\multirow{13}{*}{\begin{sideways}missing LP\end{sideways}} &
\texttt{clg-msg} & 0.275 & 0.203 & 0.241 & 0.316 & 0.207 \\
&\texttt{email} & -0.059 & -0.012 & -0.155 & -0.091 & 0.007 \\
&\texttt{math-flow} & 0.399 & 0.385 & 0.227 & 0.311 & 0.196 \\
&\texttt{ast-phys} & 0.278 & 0.391 & 0.572 & 0.373 & 0.542 \\
&\texttt{lst-fm} & --- & --- & --- & ---  & --- \\
&\texttt{forum} & 0.184 & 0.227 & 0.236 & 0.221 & 0.221 \\
&\texttt{topo} & 0.233 & 0.316 & 0.302 & 0.176 & 0.241 \\
&\texttt{act-mooc} & --- & --- & --- & --- & --- \\
&\texttt{pow-grid} & -0.144 & 0.084 & 0.173 & -0.108 & 0.189 \\
&\texttt{routers} & 0.242 & 0.005 & 0.079 & 0.285 & 0.068 \\
&\texttt{bio-yeast} & 0.216 & 0.260 & 0.362 & 0.316 & 0.302 \\
&\texttt{fb} & -0.178 & 0.450 & 0.211 & -0.159 & 0.025 \\
&\texttt{blog-cat} & 0.213 & 0.031 & 0.049 & 0.319 & 0.033 \\
\midrule
\multirow{7}{*}{\begin{sideways}future LP\end{sideways}} &
\texttt{clg-msg-t} & 0.580 & 0.589 & 0.548 & 0.578 & 0.489 \\
&\texttt{email-t} & 0.603 & 0.443 & 0.390 & 0.609 & 0.337 \\
&\texttt{math-flow-t} & 0.387 & 0.437 & 0.467 & 0.391 & 0.445 \\
&\texttt{ast-phys-t} & 0.660 & 0.705 & 0.677 & 0.508 & 0.653 \\
&\texttt{lst-fm-t} & 0.112 & -0.021 & -0.095 & 0.222 & -0.131 \\
&\texttt{forum-t} & 0.476 & 0.467 & 0.511 & 0.488 & 0.406 \\
&\texttt{topo-t} & 0.363 & 0.391 & 0.360 & 0.326 & 0.361 \\
&\texttt{act-mooc-t} & 0.439 & 0.589 & 0.573 & 0.413 & 0.594 \\
\bottomrule
\end{tabular*}
\label{tab:hop-Kend}
\end{table}

\subsection{Prediction method type}
Here we investigate if the prediction performance differ with the type of the LP methods. As all of the local-sim, global-sim and learning based methods apply on the two-hop case, we perform one way ANOVA followed by Tukey post-hoc test on these three groups of methods. As only global-sim and learning based methods apply on the three-hop case, we apply Student's t-test for this case. We summarize the results of the ANOVA for two-hop in Table~\ref{tab:pred-2-hop_Anova}, Tukey test results for two-hop in Table~\ref{tab:Pred-Tukey_Mis} and Table~\ref{tab:Pred-Tukey_Fut} respectively for missing LP and future LP, and the t-test results for the three-hop case in Table~\ref{tab:Pred-ttest}.

Table~\ref{tab:pred-2-hop_Anova} shows that the prediction performance of the prediction methods in the three groups differ significantly for almost all cases in two-hop. Post-hoc test for the missing LP in two-hop cases presented in Table~\ref{tab:Pred-Tukey_Mis} shows that across almost all datasets learning group is the clear winner, but although global-sim methods outperforms local-sim in a few datasets and evaluation methods, in most cases there are no significant difference. For the two-hop cases in future LP shown in Table~\ref{tab:Pred-Tukey_Fut}, although in \texttt{clg-msg-t}, \texttt{math-flow-t} and \texttt{math-flow-t} datasets, learning methods perform best, results are not conclusive for the rest. Table~\ref{tab:Pred-ttest} shows that, for three-hop case, there is no significant difference in performance between the global-sim and learning based methods, except for few cases like \texttt{bio-yeast}, \texttt{forum}, \texttt{blog-cat} and \texttt{lst-fm-t}. For \texttt{bio-yeast}, \texttt{forum} and \texttt{blog-cat}, global-sim methods win, and learning based methods win in \texttt{lst-fm-t}. However, the number of the test node pairs for \texttt{bio-yeast}, \texttt{forum} and \texttt{blog-cat} in three-hop case is low, so, over-fitting might have caused the performance degradation for the learning based methods. Overall, the results indicate that the relative performance of different types of LP methods vary with the distance between the end nodes. Moreover, even when the distance is same, the relative performance of the method types vary with the prediction type. The results in this section advocate for the proposed distance controlled and prediction type controlled setup, along with the consideration of the type of the methods while comparing performance of multiple methods, in order to evaluate an LP method in a robust and detailed manner.

\begin{table}[ht]
\centering
\caption{One Way ANOVA P-values for the three groups of LP methods: local-sim, global-sim and learning, on the test set of two-hop away node pairs}
\begin{tabular*}{\textwidth}{@{\extracolsep\fill}llccccc}
\toprule

\multirow{2}{*}{} & \multirow{2}{*}{DATASET} & \multirow{2}{*}{AUROC} & \multicolumn{2}{c}{AUPR} & \multicolumn{2}{c}{Pr@P} \\ \cmidrule(r){4-5} \cmidrule(r){6-7}
  &    & & balanced & imbalanced & balanced & imbalanced \\
\midrule
\multirow{15}{*}{\begin{sideways}missing LP\end{sideways}} &
        \texttt{clg-msg} & $<$.001*** & $<$.001*** & $<$.001*** & $<$.001*** & $<$.001*** \\
        &\texttt{email} & $<$.001*** & $<$.001*** & $<$.001*** & $<$.001*** & $<$.001*** \\
        &\texttt{math-flow} & $<$.001*** & $<$.001*** & $<$.001*** & $<$.001*** & $<$.001*** \\
        &\texttt{ast-phys} & 0.014* & 0.016** & 0.001** & 0.019* & 0.015** \\
        &\texttt{lst-fm} & $<$.001*** & $<$.001*** & $<$.001*** & $<$.001*** & $<$.001*** \\
       & \texttt{iacontact} & 0.136 & 0.126 & $<$.001*** & 0.508 & 0.012** \\
       & \texttt{forum} & $<$.001*** & $<$.001*** & $<$.001*** & $<$.001*** & $<$.001*** \\
       & \texttt{topo} & 0.005** & $<$.001*** & $<$.001*** & 0.028* & $<$.001*** \\
        &\texttt{act-mooc} & 0.541 & 0.449 & 0.437 & 0.687 & 0.410 \\
       & \texttt{arxiv} & 0.001** & $<$.001*** & $<$.001*** & $<$.001*** & $<$.001*** \\
       & \texttt{pow-grid} & 0.415 & 0.003** & $<$.001*** & 0.012* & 0.008** \\
        &\texttt{routers} & $<$.001*** & $<$.001*** & $<$.001*** & $<$.001*** & $<$.001*** \\
       & \texttt{bio-yeast} & $<$.001*** & $<$.001*** & $<$.001*** & $<$.001*** & $<$.001*** \\
        &\texttt{fb} & 0.070 & 0.011* & $<$.001*** & 0.021* & 0.003** \\
       & \texttt{blog-cat} & $<$.001*** & $<$.001*** & $<$.001*** & $<$.001*** & $<$.001*** \\
        \midrule
        \multirow{9}{*}{\begin{sideways}future LP\end{sideways}}
       & \texttt{clg-msg-t} & $<$.001*** & $<$.001*** & $<$.001*** & $<$.001*** & $<$.001*** \\
       & \texttt{email-t} & 0.161 & 0.012* & $<$.001*** & 0.157 & $<$.001*** \\
       & \texttt{math-flow-t} & $<$.001*** & $<$.001*** & $<$.001*** & $<$.001*** & $<$.001*** \\
       & \texttt{ast-phys-t} & 0.007** & $<$.001*** & 0.003** & 0.104 & $<$.001*** \\
       & \texttt{lst-fm-t} & 0.001** & 0.002** & 0.001** & $<$.001*** & 0.002** \\
        &\texttt{iacontact-t} & 0.096 & 0.008** & $<$.001*** & 0.162 & $<$.001*** \\
       & \texttt{forum-t} & $<$.001*** & 0.003** & $<$.001*** & $<$.001*** & 0.027* \\
       & \texttt{topo-t} & 0.085 & 0.042* & $<$.001*** & 0.046* & $<$.001*** \\
       & \texttt{act-mooc-t} & $<$.001*** & $<$.001*** & $<$.001*** & $<$.001*** & $<$.001*** \\
        \bottomrule
    \end{tabular*}

    \label{tab:pred-2-hop_Anova}
\end{table}

\begin{landscape}
\begin{table*}[ht]
\centering
\caption{Tukey's post-hoc test results among the three groups of LP methods: local-sim, global-sim and learning, on the test set of two-hop away node pairs for missing LP}
\begin{adjustbox}{width=1\textwidth,center}
\begin{tabular}{llclclclclcl}
\toprule
\multirow{3}{*}{\textbf{DATASET}} & \multirow{3}{*}{\textbf{METHOD GROUP}} & \multicolumn{2}{c}{\multirow{3}{*}{\textbf{AUROC}}} & \multicolumn{4}{c}{\textbf{AUPR}} & \multicolumn{4}{c}{\textbf{Pr@P}} \\ 
\cmidrule(r){5-8} \cmidrule(r){9-12}
 & & \multicolumn{2}{c}{} & \multicolumn{2}{c}{\textbf{balanced}} & \multicolumn{2}{c}{\textbf{imbalanced}} & \multicolumn{2}{c}{\textbf{balanced}} & \multicolumn{2}{c}{\textbf{imbalanced}} \\ 
\cmidrule(r){3-4} \cmidrule(r){5-6} \cmidrule(r){7-8} \cmidrule(r){9-10} \cmidrule(r){11-12}
 & &\textbf{P-Value} & \textbf{Winner} & \textbf{P-Value} & \textbf{Winner} & \textbf{P-Value} & \textbf{Winner} & \textbf{P-Value} & \textbf{Winner} & \textbf{P-Value} & \textbf{Winner} \\ 
\midrule
\texttt{clg-msg} & learning vs global-sim & $<$.001*** & learning & $<$.001*** & learning & $<$.001*** & learning & $<$.001*** & learning & $<$.001*** & learning \\
 & global-sim vs local-sim & 0.762 & --- & 0.787 & --- & 0.631 & --- & 0.590 & --- & 0.415 & --- \\
 & learning vs local-sim & $<$.001*** & learning & $<$.001*** & learning & $<$.001*** & learning & $<$.001*** & learning & $<$.001*** & learning \\
\midrule
\texttt{email} & learning vs global-sim & 0.006** & learning & 0.004** & learning & 0.001** & learning & 0.001** & learning & 0.001** & learning \\
 & global-sim vs local-sim & 0.086 & --- & 0.022* & global-sim & 0.031* & global-sim & $<$.001*** & global-sim & 0.006** & global-sim \\
 & learning vs local-sim & $<$.001*** & learning & $<$.001*** & learning & $<$.001*** & learning & $<$.001*** & learning & $<$.001*** & learning \\
\midrule
\texttt{math-flow} 
 & learning vs global-sim & $<$.001*** & learning & $<$.001*** & learning & $<$.001*** & learning & $<$.001*** & learning & $<$.001*** & learning \\
 & global-sim vs local-sim & 0.459 & --- & 0.518 & --- & 0.297 & --- & 0.116 & --- & 0.353 & --- \\
 & learning vs local-sim & $<$.001*** & learning & $<$.001*** & learning & $<$.001*** & learning & $<$.001*** & learning & $<$.001*** & learning \\
\midrule
\texttt{ast-phys}
 & learning vs global-sim & 0.221 & --- & 0.251 & --- & 0.181 & --- & 0.049* & learning & 0.178 & --- \\
 & global-sim vs local-sim & 0.767 & --- & 0.738 & --- & 0.894 & --- & 0.804 & --- & 0.862 & --- \\
  & learning vs local-sim & 0.029* & learning & 0.030* & learning & 0.050* & learning & 0.005** & learning & 0.039** & learning \\
\midrule
\texttt{lst-fm} 
 & learning vs global-sim & $<$.001*** & learning & 0.007** & learning & $<$.001*** & learning & 0.005** & learning & 0.572 & --- \\
 & global-sim vs local-sim & 0.267 & --- & 0.045* & global-sim & 0.296 & --- & 0.041* & global-sim & 0.020** & global-sim \\
 & learning vs local-sim & $<$.001*** & learning & $<$.001*** & learning & $<$.001*** & learning & $<$.001*** & learning & $<$.001*** & learning \\
\midrule
\texttt{iacontact}
 & learning vs global-sim & 0.452 & --- & 0.606 & --- & 0.507 & --- & 0.376 & --- & 0.924 & --- \\
 & global-sim vs local-sim & 0.117 & --- & 0.123 & --- & 0.568 & --- & 0.008** & global-sim & 0.040** & global-sim \\
  & learning vs local-sim & 0.272 & --- & 0.186 & --- & 0.958 & --- & $<$.001*** & learning & 0.010** & learning \\
\midrule
\texttt{forum}
 & learning vs global-sim & $<$.001*** & learning & $<$.001*** & learning & $<$.001*** & learning & $<$.001*** & learning & $<$.001*** & learning \\
 & global-sim vs local-sim & 0.217 & --- & 0.895 & --- & 0.208 & --- & 0.630 & --- & 0.674 & --- \\
  & learning vs local-sim & 0.007** & learning & $<$.001*** & learning & 0.038* & learning & $<$.001*** & learning & $<$.001*** & learning \\
\midrule
\texttt{topo}
 & learning vs global-sim & 0.197 & --- & 0.050* & learning & 0.481 & --- & 0.001** & learning & 0.005** & learning \\
 & global-sim vs local-sim & 0.602 & --- & 0.656 & --- & 0.558 & --- & 0.188 & --- & 0.472 & --- \\
  & learning vs local-sim & 0.009** & learning & 0.002** & learning & 0.0336* & learning & $<$.001*** & learning & $<$.001*** & learning \\
\midrule
\texttt{act-mooc}
 & learning vs global-sim & 0.624 & --- & 0.545 & --- & 0.876 & --- & 0.878 & --- & 0.830 & --- \\
 & global-sim vs local-sim & 0.980 & --- & 0.977 & --- & 0.981 & --- & 0.851 & --- & 0.879 & --- \\
  & learning vs local-sim & 0.782 & --- & 0.721 & --- & 0.738 & --- & 0.441 & --- & 0.432 & --- \\
\midrule
\texttt{arxiv} 
 & learning vs global-sim & 0.221 & --- & 0.251 & --- & 0.181 & --- & 0.999 & --- & 0.980 & --- \\
 & global-sim vs local-sim & 0.767 & --- & 0.738 & --- & 0.894 & --- & $<$.001*** & global-sim & 0.001** & global-sim \\
& learning vs local-sim & 0.029* & learning & 0.030* & learning & 0.050* & learning & $<$.001*** & learning & $<$.001*** & learning \\
\midrule
\texttt{pow-grid} 
 & learning vs global-sim & 0.562 & --- & 0.148 & --- & 0.930 & --- & 0.040* & learning & 0.015** & learning \\
 & global-sim vs local-sim & 0.996 & --- & 0.626 & --- & 0.041* & global-sim & 0.004** & global-sim & 0.666 & --- \\
 & learning vs local-sim & 0.635 & --- & 0.007** & learning & 0.009** & learning & $<$.001*** & learning & 0.212 & --- \\
\midrule
 
\texttt{routers} 
 & learning vs global-sim & 0.059 & --- & 0.008** & learning & 0.057 & --- & $<$.001*** & learning & $<$.001*** & learning \\
 & global-sim vs local-sim & 0.331 & --- & 0.054 & --- & 0.707 & --- & $<$.001*** & global-sim & 0.002** & global-sim \\
 & learning vs local-sim & $<$.001*** & learning & $<$.001*** & learning & 0.003** & learning & $<$.001*** & learning & $<$.001*** & learning \\
\midrule
\texttt{bio-yeast}
 & learning vs global-sim & 0.004** & learning & 0.017* & learning & 0.002** & learning & 0.004** & learning & 0.006** & learning \\
 & global-sim vs local-sim & 0.692 & --- & 0.600 & --- & 0.494 & --- & 0.130 & --- & 0.194 & --- \\
  & learning vs local-sim & $<$.001*** & learning & $<$.001*** & learning & $<$.001*** & learning & $<$.001*** & learning & $<$.001*** & learning \\
\midrule
\texttt{fb}
 & learning vs global-sim & 0.657 & --- & 0.414 & --- & 0.633 & --- & 0.173 & --- & 0.335 & --- \\
 & global-sim vs local-sim & 0.582 & --- & 0.456 & --- & 0.368 & --- & 0.167 & --- & 0.334 & --- \\
  & learning vs local-sim & 0.073 & --- & 0.014* & learning & 0.020* & learning & $<$.001*** & learning & 0.004** & learning \\
\midrule
\texttt{blog-cat}
 & learning vs global-sim & 0.002** & learning & 0.001** & learning & 0.001** & learning & $<$.001*** & learning & $<$.001*** & learning \\
 & global-sim vs local-sim & 0.774 & --- & 0.816 & --- & 0.680 & --- & 0.546 & --- & 0.712 & --- \\
  & learning vs local-sim & $<$.001*** & learning & $<$.001*** & learning & $<$.001*** & learning & $<$.001*** & learning & $<$.001*** & learning \\
\bottomrule
\end{tabular}
\end{adjustbox}
\label{tab:Pred-Tukey_Mis}
\end{table*}
\end{landscape}

\begin{table*}[h]
\centering
\caption{Tukey's post-hoc test results among the three groups of LP methods: local-sim, global-sim and learning, on the test set of two-hop away node pairs for future LP}
\begin{adjustbox}{width=1\textwidth,center}
\begin{tabular}{llclclclclcl}
\toprule
\multirow{3}{*}{\textbf{DATASET}} & \multirow{3}{*}{\textbf{METHOD GROUP}} & \multicolumn{2}{c}{\multirow{3}{*}{\textbf{AUROC}}} & \multicolumn{4}{c}{\textbf{AUPR}} & \multicolumn{4}{c}{\textbf{Pr@P}} \\ 
\cmidrule(r){5-8} \cmidrule(r){9-12}
 & & \multicolumn{2}{c}{} & \multicolumn{2}{c}{\textbf{balanced}} & \multicolumn{2}{c}{\textbf{imbalanced}} & \multicolumn{2}{c}{\textbf{balanced}} & \multicolumn{2}{c}{\textbf{imbalanced}} \\ 
\cmidrule(r){3-4} \cmidrule(r){5-6} \cmidrule(r){7-8} \cmidrule(r){9-10} \cmidrule(r){11-12}
 & &\textbf{P-Value} & \textbf{Winner} & \textbf{P-Value} & \textbf{Winner} & \textbf{P-Value} & \textbf{Winner} & \textbf{P-Value} & \textbf{Winner} & \textbf{P-Value} & \textbf{Winner} \\ 
\midrule
\texttt{clg-msg-t}  & learning vs global-sim & 0.001** & learning & 0.007** & learning & $<$.001*** & learning & $<$.001*** & learning & 0.001** & learning \\
 & global-sim vs local-sim & 0.366 & --- & 0.933 & --- & 0.785 & --- & 0.109 & --- & 0.706 & --- \\
  & learning vs local-sim & 0.159 & --- & 0.026* & learning & 0.002** & learning & 0.299 & --- & 0.027* & learning \\ \midrule
\texttt{email-t} 
 & learning vs global-sim & 0.970 & --- & 0.879 & --- & 0.715 & --- & 0.977 & --- & 0.801 & --- \\
 & global-sim vs local-sim & 0.432 & --- & 0.149 & --- & $<$.001*** & global-sim & 0.414 & --- & 0.004** & global-sim \\ 
 & learning vs local-sim & 0.137 & --- & 0.009** & learning & $<$.001*** & learning & 0.133 & --- & $<$.001*** & learning \\ \midrule
\texttt{math-flow-t}  & learning vs global-sim & 0.001** & learning & $<$.001*** & learning & $<$.001*** & learning & $<$.001*** & learning & $<$.001*** & learning \\
 & global-sim vs local-sim & 0.830 & --- & 0.764 & --- & 0.204 & --- & 0.787 & --- & 0.521 & --- \\
 & learning vs local-sim & $<$.001*** & learning & $<$.001*** & learning & $<$.001*** & learning & $<$.001*** & learning & $<$.001*** & learning \\ \midrule
\texttt{ast-phys-t}  & learning vs global-sim & 0.037* & global-sim & 0.004** & global-sim & 0.072 & --- & 0.173 & --- & 0.011* & global-sim \\
 & global-sim vs local-sim & 0.974 & --- & 1.000 & --- & 0.873 & --- & 0.922 & --- & 0.615 & --- \\
 & learning vs local-sim & 0.074 & --- & 0.005** & local-sim & 0.014* & local-sim & 0.404 & --- & $<$.001*** & local-sim \\ \midrule
\texttt{lst-fm-t}  & learning vs global-sim & 0.852 & --- & 0.772 & --- & 0.957 & --- & 0.433 & --- & 0.083 & --- \\ 
 & global-sim vs local-sim & 0.029* & global-sim & 0.008** & global-sim & 0.010* & global-sim & 0.063 & --- & 0.002** & global-sim \\
 & learning vs local-sim & $<$.001*** & learning & 0.002** & learning & $<$.001*** & learning & $<$.001*** & learning & 0.015* & learning \\ \midrule
\texttt{iacontact-t} & learning vs global-sim & 0.911 & --- & 0.568 & --- & 0.928 & --- & 0.935 & --- & 0.222 & --- \\
 & global-sim vs local-sim & 0.159 & --- & 0.013* & global-sim & $<$.001*** & global-sim & 0.487 & --- & $<$.001*** & global-sim \\ 
 & learning vs local-sim & 0.090 & --- & 0.011* & learning & $<$.001*** & learning & 0.141 & --- & $<$.001*** & learning \\ \midrule
\texttt{forum-t}  & learning vs global-sim & 0.001** & learning & 0.018* & learning & 0.006** & learning & $<$.001*** & learning & 0.058 & --- \\
 & global-sim vs local-sim & 0.265 & --- & 0.962 & --- & 0.992 & --- & 0.003** & local-sim & 0.858 & --- \\
 & learning vs local-sim & 0.167 & --- & 0.045* & learning & 0.009** & learning & 0.744 & --- & 0.244 & --- \\  \midrule
\texttt{topo-t}  & learning vs global-sim & 0.660 & --- & 0.516 & --- & 0.200 & --- & 0.521 & --- & 0.157 & --- \\
 & global-sim vs local-sim & 0.625 & --- & 0.607 & --- & 0.277 & --- & 0.622 & --- & 0.366 & --- \\ 
 & learning vs local-sim & 0.091 & --- & 0.050* & learning & $<$.001*** & learning & 0.055 & --- & $<$.001*** & learning \\ \midrule
\texttt{act-mooc-t} 
 & learning vs global-sim & 0.003** & global-sim & $<$.001*** & global-sim & 0.086 & --- & 0.024* & global-sim & 0.079 & --- \\
 & global-sim vs local-sim & 0.939 & --- & 0.859 & --- & 0.583 & --- & 0.825 & --- & 0.468 & --- \\
 & learning vs local-sim & 0.001** & local-sim & $<$.001*** & local-sim & 0.003** & local-sim & 0.003** & local-sim & $<$.001*** & local-sim \\
\bottomrule
\end{tabular}
\end{adjustbox}
\label{tab:Pred-Tukey_Fut}
\end{table*}

\begin{table*}[ht]
    \centering
    \caption{t-test results between the two groups of LP methods: global-sim and learning, on the test set of three-hop away node pairs}
\begin{adjustbox}{width=1\textwidth}

    \begin{tabular}{llclclclclcl}
        \toprule
       \multirow{3}{*}{} & \multirow{3}{*}{\textbf{DATASET}} & \multicolumn{2}{c}{\multirow{3}{*}{\textbf{AUROC}}} & \multicolumn{4}{c}{\textbf{AUPR}} & \multicolumn{4}{c}{\textbf{Pr@P}} \\ 
\cmidrule(r){5-8} \cmidrule(r){9-12}

 &  & \multicolumn{2}{c}{} & \multicolumn{2}{c}{\textbf{balanced}} & \multicolumn{2}{c}{\textbf{imbalanced}} & \multicolumn{2}{c}{\textbf{balanced}} & \multicolumn{2}{c}{\textbf{imbalanced}} \\ 
\cmidrule(r){3-4} \cmidrule(r){5-6} \cmidrule(r){7-8} \cmidrule(r){9-10} \cmidrule(r){11-12}
 &  & \textbf{P-Value} & \textbf{Winner} & \textbf{P-Value} & \textbf{Winner} & \textbf{P-Value} & \textbf{Winner} & \textbf{P-Value} & \textbf{Winner} & \textbf{P-Value} & \textbf{Winner} \\
        \midrule
      \multirow{15}{*}{\begin{sideways}missing LP\end{sideways}}
       & \texttt{clg-msg} & 0.043* & global-sim & 0.054 & --- & 0.082 & --- & 0.043* & global-sim & 0.057 & --- \\
        & \texttt{email} & 0.039* & global-sim & 0.157 & --- & 0.078 & --- & \ $<$.001*** & global-sim & 0.532 & --- \\
        & \texttt{math-flow} & 0.365 & --- & 0.206 & --- & 0.195 & --- & 0.462 & --- & 0.199 & --- \\
       & \texttt{ast-phys} & 0.329 & --- & 0.479 & --- & 0.303 & --- & 0.232 & --- & 0.388 & --- \\
        & \texttt{lst-fm} & --- & --- & --- & --- & --- & --- & --- & --- & --- & --- \\
        & \texttt{iacontact} & --- & --- & --- & --- & --- & --- & --- & --- & --- & --- \\
      &  \texttt{forum} & 0.020* & global-sim & 0.039* & global-sim & 0.048* & global-sim & 0.016* & global-sim & 0.036* & global-sim \\
       &  \texttt{topo} & 0.558 & --- & 0.173 & --- & 0.215 & --- & 0.694 & --- & 0.662 & --- \\
       &  \texttt{act-mooc} & --- & --- & --- & --- & --- & --- & --- & --- & --- & --- \\
        & \texttt{arxiv} & --- & --- & --- & --- & --- & --- & --- & --- & --- & --- \\
        & \texttt{pow-grid} & 0.367 & --- & 0.287 & --- & 0.164 & --- & 0.581 & --- & 0.251 & --- \\
       &  \texttt{routers} & 0.458 & --- & 0.325 & --- & 0.470 & --- & 0.391 & --- & 0.463 & --- \\
        & \texttt{bio-yeast} & \ $<$.001*** & global-sim & 0.019* & global-sim & 0.012* & global-sim & \ $<$.001*** & global-sim & 0.003* & global-sim \\
       &  \texttt{fb} & 0.328 & --- & 0.650 & --- & 0.552 & --- & \ $<$.001*** & learning & 0.925 & --- \\
        & \texttt{blog-cat} & 0.007** & global-sim & 0.004** & global-sim & 0.003** & global-sim & 0.094 & --- & \ $<$.001*** & global-sim \\
        \midrule
       \multirow{8}{*}{\begin{sideways}future LP\end{sideways}}
       &  \texttt{clg-msg-t} & 0.288 & --- & 0.328 & --- & 0.227 & --- & 0.293 & --- & 0.150 & --- \\
        & \texttt{email-t} & 0.611 & --- & 0.599 & --- & 0.165 & --- & 0.400 & --- & 0.241 & --- \\
        & \texttt{math-flow-t} & 0.516 & --- & 0.356 & --- & 0.297 & --- & 0.528 & --- & 0.287 & --- \\
        & \texttt{ast-phys-t} & 0.705 & --- & 0.554 & --- & 0.527 & --- & 0.730 & --- & 0.545 & --- \\
        & \texttt{lst-fm} & 0.017* & learning & 0.473 & --- & 0.256 & --- & 0.010** & learning & 0.034* & learning \\
        & \texttt{iacontact} & --- & --- & --- & --- & --- & --- & --- & --- & --- & --- \\
       &  \texttt{forum-t} & 0.051 & --- & 0.052 & --- & 0.117 & --- & 0.044* & global-sim & 0.073 & --- \\
       &  \texttt{topo-t} & 0.616 & --- & 0.248 & --- & 0.108 & --- & 0.606 & --- & 0.337 & --- \\
        & \texttt{act-mooc-t} & 0.613 & --- & 0.717 & --- & 0.898 & --- & 0.785 & --- & 0.998 & --- \\
        \bottomrule
    \end{tabular}
    \end{adjustbox}
    \label{tab:Pred-ttest}
\end{table*}

\subsection{Network type}
As most of the LP methods apply on undirected networks, while applying those methods on directed networks, they are converted to undirected ones ignoring directions. Here we investigate whether this information loss affects their prediction performance. For each of future and missing LP, we divide the datasets into two groups: directed and undirected, and perform Student's t-test for a LP method given its accuracy against a particular evaluation metric. To avoid getting overloaded with the high number of the LP methods, we conduct the test once for each type of methods: local-sim, global-sim and learning, by considering the performance score of the best performing method in that group. In Table~\ref{tab:Net-Dir}, we report the test results. For the best performing methods in each group given an evaluation metric, readers may refer to Appendix A.

We find few tests achieve the significance level, and for all such cases the undirected networks are the winner. More interestingly, the undirected version wins for all cases for the future LP when the evaluation method is the imbalanced version of AUPR or $Pr@P$. This evaluation setup is the most prominent: we are considering future LP on a longitudinal network, and our test set is imbalanced. The results in this particular setup is indicative for the hypothesis that the information loss in directed networks due to ignoring directions may adversely affect the prediction performance. Note that, we do not strike out the possibility of the presence of any confounder inherent to the directed networks which may result the performance degradation. Study on such confounders is out of the scope of this paper.

\begin{table*}[ht]
\centering
\caption{t-test results for LP in directed vs. undirected networks}
\begin{adjustbox}{width=1\textwidth,center}

\begin{tabular}{llclclclclcl}
\toprule
\multirow{3}{*}{\textbf{DATASET}} & \multirow{3}{*}{\textbf{\makecell{METHOD\\TYPE}}} & \multicolumn{2}{c}{\multirow{3}{*}{\textbf{AUROC}}} & \multicolumn{4}{c}{\textbf{AUPR}} & \multicolumn{4}{c}{\textbf{Pr@P}} \\ 
\cmidrule(r){5-8} \cmidrule(r){9-12}
 & & \multicolumn{2}{c}{} & \multicolumn{2}{c}{\textbf{balanced}} & \multicolumn{2}{c}{\textbf{imbalanced}} & \multicolumn{2}{c}{\textbf{balanced}} & \multicolumn{2}{c}{\textbf{imbalanced}} \\ 
\cmidrule(r){3-4} \cmidrule(r){5-6} \cmidrule(r){7-8} \cmidrule(r){9-10} \cmidrule(r){11-12}
 & &\textbf{P-Value} & \textbf{Winner} & \textbf{P-Value} & \textbf{Winner} & \textbf{P-Value} & \textbf{Winner} & \textbf{P-Value} & \textbf{Winner} & \textbf{P-Value} & \textbf{Winner} \\
\midrule
\multirow{3}{*}{future two-hop} & local-sim & 0.736 & --- & 0.880 & --- & 0.025* & undirected & 0.687 & --- & 0.018* & undirected \\
 & global-sim & 0.750 & --- & 0.787 & --- & 0.036* & undirected & 0.770 & --- & 0.045* & undirected \\
 & learning & 0.737 & --- & 0.688 & --- & 0.038* & undirected & 0.681 & --- & 0.045* & undirected \\
\midrule
\multirow{3}{*}{missing two-hop} & local-sim & 0.199 & --- & 0.091 & --- & 0.030* & undirected & 0.152 & --- & 0.072 & --- \\
 & global-sim & 0.577 & --- & 0.433 & --- & 0.607 & --- & 0.512 & --- & 0.590 & --- \\
 & learning & 0.384 & --- & 0.322 & --- & 0.566 & --- & 0.432 & --- & 0.541 & --- \\
\midrule
\multirow{2}{*}{future three-hop} & global-sim & 0.716 & --- & 0.757 & --- & 0.593 & --- & 0.049* & undirected & 0.593 & --- \\
 & learning & 0.343 & --- & 0.376 & --- & 0.416 & --- & 0.068 & --- & 0.416 & --- \\
\midrule
\multirow{2}{*}{missing three-hop} & global-sim & 0.505 & --- & 0.381 & --- & 0.119 & --- & 0.197 & --- & 0.114 & --- \\
 & learning & 0.020* & undirected & 0.062 & --- & 0.062 & --- & 0.173 & --- & 0.093 & --- \\
\bottomrule
\end{tabular}
\end{adjustbox}
\label{tab:Net-Dir}
\end{table*}

\subsection{Class imbalance and Early retrieval}
In this section, we analyze how class imbalance affects the evaluation of LP. First we investigate if the relative performance of the LP methods differ subject to AUROC vs. AUPR. Both of AUROC and AUPR consider all threshold values, but AUROC is not affected by class imbalance whereas AUPR is. We consider both balanced as well as imbalanced (skew level $1:10$) test set for AUPR. Next, we take the imbalance sensitive measure AUPR, and compare the relative performance of the LP methods on it for the balanced and imbalanced version of the test set.
We further do the same with the single threshold imbalance sensitive metric $Pr@P$. We consider Kendall?s Tau Rank Correlation Coefficient ($\tau$) to measure the (in)consistently of ranks of the LP methods when evaluated in each pair of setups stated above. The $\tau$ values for each of the datasets are shown in Table~\ref{tab:Cls-imb-P}. We summarize our observations as follows.

Considering the correlation values for AUROC vs. AUPR, it is evident that for almost every dataset across all distances (two-hop/three-hop) and prediction types (future LP/missing LP), the correlation decreases when the imbalanced version of the test set is considered over the balanced one. For a significant number of cases (such as, in two-hop scenario, \texttt{iacontact}, \texttt{forum}, \texttt{arxiv}, \texttt{pow-grid}, etc.) the correlation in imbalance scenario are considerably low. This indicates that when tested on imbalance dataset, there are significant disagreements on the performance of various LP methods when AUPR is used against AUROC. The correlation values for AUPR: balanced vs. imbalanced, and $Pr@P$: balanced vs. imbalanced indicate some disagreements, which is higher for $Pr@P$. This observation is consistent across distances and prediction types. These results show that if LP methods are tested on balanced test set, the conclusions may be erroneous, because they may behave differently in an imbalance setup which is inherent to the LP problem. It also suggests that although AUROC has been traditionally used to evaluate LP methods, their performance significantly vary if skew sensitive metrics like AUPR or $Pr@P$ are applied on an imbalanced test setup. Hence we recommend AUPR or $Pr@P$ as a single point summary metric applied on an imbalanced setup to evaluate LP, while maintaining other factor as described in Section~\ref{exp-setup-gt}. Unlike $Pr@P$, as AUPR covers all threshold values, it can capture the whole spectrum of the precision-recall space, it should be the first choice.
\subsubsection{Early Retrieval:}
For certain applications like recommender systems, LP methods must perform well at the early retrieval phase, and it becomes critical when the data is highly skewed. Here we investigate how class imbalance affects early retrieval. The ROC curve, rather the Concentrated ROC (CROC) curve~\cite{swamidass2010croc} can reflect the performance in the early retrieval stage. However, as it is skew insensitive, it is unable to reflect the performance in imbalanced setup. Saito et al.~\cite{saito2015precision} has shown how PRC can be used to understand the effect of imbalance in early retrieval. We first investigate how the performance on the single point metric $Pr@P/2$ (both balanced and imbalanced versions) correlate with that of AUROC, and the balanced and imbalanced versions of $Pr@P/2$ correlate with each other, and then show an example ROC plot demonstrating the different behavior of the prediction methods in the two test settings. For each dataset, we measure the Kendall?s Tau Rank Correlation Coefficient ($\tau$) of the LP methods for the pair of cases stated above in our controlled setup.

\begin{figure}[h]
\centering
\includegraphics[width=0.9\textwidth]{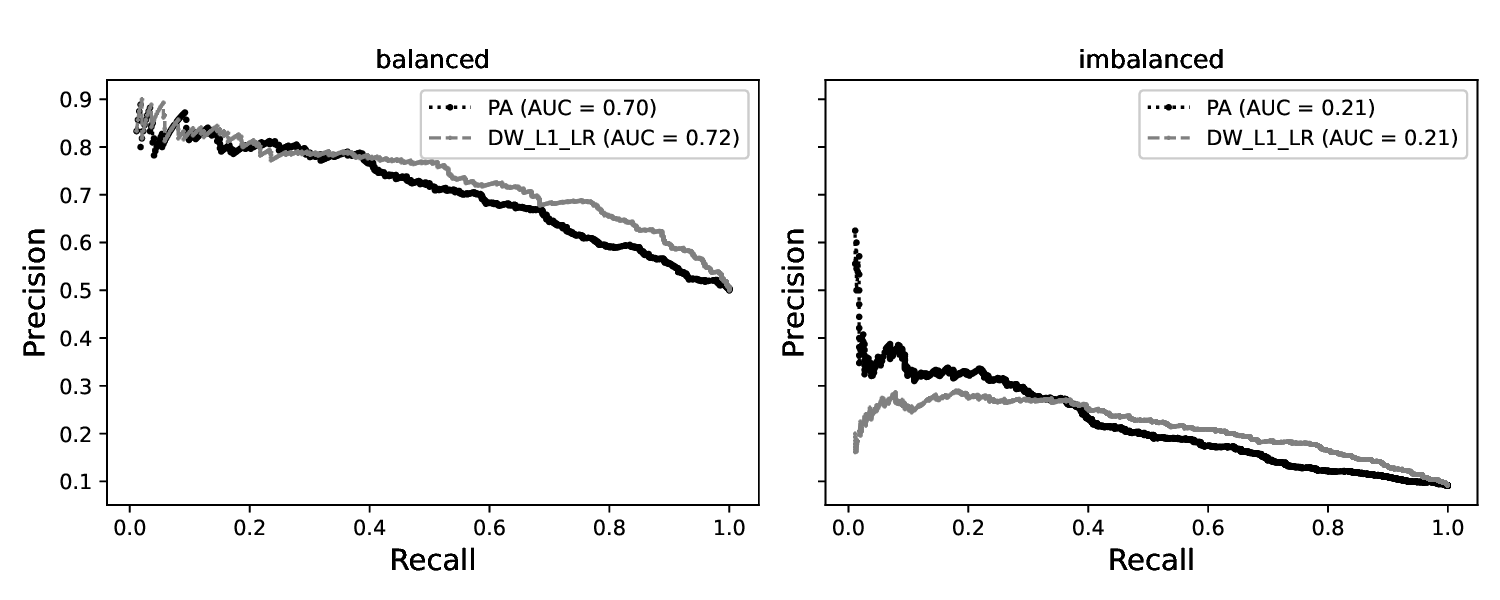}
 \caption{Effect of class imbalance on early retrieval: PR curves for $PA$ and \textit{DW\_L1\_LR} for the \texttt{email} dataset in two-hop test scenario}\label{fig:empil}
 \end{figure}

Table~\ref{tab:Cls-imb-P/2} shows that the correlations with AUROC are significantly decreasing from $Pr@P/2$-balanced to $Pr@P/2$-imbalanced case, which implies that with increasing imbalance, AUROC becomes very inconsistent with the early retrieval behavior. In many cases, the correlation between the balanced and the imbalanced version of $Pr@P/2$ is considerably low, which indicates that, in early retrieval phase if we consider imbalanced test set, LP methods behave differently as that of the balanced case. So, we recommend carefully investigating the early retrieval performance of the prediction methods considering imbalanced test set in the controlled setup presented in this paper, using proper visualization technique like PR curve. Figure~\ref{fig:empil} presents a case where the imbalance affect the early retrieval. It shows the PRC plots for the two methods $PA$ and \textit{DW\_L1\_LR} for the \texttt{email} dataset in two-hop test scenario, both for balanced and imbalanced cases. It shows that although they perform comparably in both of the balanced and imbalanced cases, $PA$ outperforms \textit{DW\_L1\_LR} clearly in the early retrieval phase in the imbalanced case.


 \begin{table*}[h]
\centering
\caption{Kendall’s $\tau$ correlation results for pairs of selected evaluation measures to understand the effect of class imbalance}
\adjustbox{width=1\textwidth,center}{%
\begin{tabular}{llcccccccc}
\toprule
 & \multirow{6}{*}{\textbf{DATASET}} & \multicolumn{4}{c}{\textbf{two-hop}} & \multicolumn{4}{c}{\textbf{three-hop}} \\
\cmidrule(lr){3-6} \cmidrule(lr){7-10}
& & \textbf{\makecell{AUROC\\ vs}} & \textbf{\makecell{AUROC\\ vs}} & \textbf{\makecell{AUPR\\(balanced)\\ vs}} & \textbf{\makecell{Pr@P\\(balanced)\\ vs}} & \textbf{\makecell{AUROC\\ vs}} & \textbf{\makecell{AUROC\\ vs}} & \textbf{\makecell{AUPR\\(balanced)\\ vs}} & \textbf{\makecell{Pr@P\\(balanced)\\ vs}} \\
& & \textbf{\makecell{AUPR\\(balanced)}} & \textbf{\makecell{AUPR\\(imbalanced)}} & \textbf{\makecell{AUPR\\(imbalanced)}} & \textbf{\makecell{Pr@P\\(imbalanced)}} & \textbf{\makecell{AUPR\\(balanced)}} & \textbf{\makecell{AUPR\\(imbalanced)}} & \textbf{\makecell{AUPR\\(imbalanced)}} & \textbf{\makecell{Pr@P\\(imbalanced)}} \\
\midrule
\multirow{17}{*}{\rotatebox[origin=c]{90}{missing LP}}
& \texttt{clg-msg} & 0.81 & 0.72 & 0.75 & 0.54 & 0.73 & 0.46 & 0.58 & 0.39 \\
& \texttt{email} & 0.86 & 0.75 & 0.76 & 0.56 & --- & --- & --- & --- \\
& \texttt{math-flow} & 0.90 & 0.79 & 0.87 & 0.71 & 0.87 & 0.78 & 0.85 & 0.53 \\
& \texttt{ast-phys} & 0.89 & 0.84 & 0.93 & 0.85 & 0.17 & 0.17 & 0.66 & 0.47 \\
& \texttt{lst-fm} & 0.81 & 0.75 & 0.84 & 0.61 & --- & --- & --- & --- \\
& \texttt{iacontact} & 0.73 & 0.48 & 0.56 & 0.21 & --- & --- & --- & --- \\
& \texttt{forum} & 0.80 & 0.65 & 0.72 & 0.45 & 0.81 & 0.67 & 0.70 & 0.37 \\
& \texttt{topo} & 0.85 & 0.81 & 0.95 & 0.64 & 0.79 & 0.59 & 0.72 & 0.58 \\
& \texttt{act-mooc} & 0.95 & 0.91 & 0.95 & 0.87 & --- & --- & --- & --- \\

& \texttt{arxiv} & 0.77 & 0.54 & 0.71 & 0.56 & --- & --- & --- & --- \\
& \texttt{pow-grid} & 0.65 & 0.52 & 0.71 & -0.14 & 0.79 & 0.56 & 0.67 & 0.30 \\
& \texttt{routers} & 0.74 & 0.71 & 0.80 & 0.51 & 0.77 & 0.55 & 0.67 & 0.51 \\
& \texttt{bio-yeast} & 0.86 & 0.78 & 0.89 & 0.77 & 0.49 & 0.38 & 0.59 & 0.09 \\
& \texttt{fb} & 0.90 & 0.80 & 0.87 & 0.73 & --- & --- & --- & --- \\
& \texttt{blog-cat} & 0.89 & 0.83 & 0.91 & 0.83 & 0.70 & 0.60 & 0.64 & 0.45 \\

\midrule
\multirow{8}{*}{\rotatebox[origin=c]{90}{future LP}}
& \texttt{clg-msg-t} & 0.84 & 0.82 & 0.84 & 0.66 & 0.85 & 0.84 & 0.86 & 0.73 \\
& \texttt{email-t} & 0.76 & 0.67 & 0.82 & 0.51 & 0.87 & 0.71 & 0.73 & 0.59 \\
& \texttt{math-flow-t} & 0.93 & 0.88 & 0.94 & 0.83 & 0.91 & 0.84 & 0.90 & 0.84 \\
& \texttt{ast-phys-t} & 0.90 & 0.77 & 0.87 & 0.78 & 0.86 & 0.83 & 0.85 & 0.58 \\
& \texttt{lst-fm-t} & 0.82 & 0.80 & 0.93 & 0.66 & 0.35 & 0.58 & 0.68 & 0.16 \\
& \texttt{iacontact-t} & 0.82 & 0.80 & 0.78 & 0.56 & --- & --- & --- & --- \\
& \texttt{forum-t} & 0.83 & 0.79 & 0.75 & 0.54 & 0.89 & 0.84 & 0.84 & 0.72 \\
& \texttt{topo-t} & 0.91 & 0.85 & 0.93 & 0.85 & 0.91 & 0.76 & 0.82 & 0.76 \\
& \texttt{act-mooc-t} & 0.93 & 0.87 & 0.93 & 0.87 & 0.57 & 0.74 & 0.81 & 0.39 \\

\bottomrule
\end{tabular}%
}
\label{tab:Cls-imb-P}
\end{table*}


\begin{table*}[h]
\centering
\caption{Kendall’s $\tau$ correlation results for pairs of selected evaluation measures to understand the effect of class imbalance in early retrieval scenario}
\begin{adjustbox}{width=1.0\textwidth,center}

\begin{tabular}{llcccccc}
\toprule
\multirow{6}{*}{} & \multirow{6}{*}{\textbf{DATASET}} & \multicolumn{3}{c}{\textbf{two-hop}} & \multicolumn{3}{c}{\textbf{three-hop}} \\
\cmidrule(lr){3-5} \cmidrule(lr){6-8}
& & {\textbf{\makecell{AUROC\\ vs\\ Pr@P/2\\(balanced)}}} & {\textbf{\makecell{AUROC\\ vs \\Pr@P/2\\(imbalanced)}}} & {\textbf{\makecell{Pr@P/2\\(balanced)\\ vs \\Pr@P/2\\(imbalanced)}}} & {\textbf{\makecell{AUROC\\ vs\\Pr@P/2\\(balanced)}}} & {\textbf{\makecell{AUROC\\ vs \\Pr@P/2\\(imbalanced)}}} & {\textbf{\makecell{Pr@P/2\\(balanced) \\vs\\ Pr@P/2\\(imbalanced)}}} \\
\midrule
\multirow{13}{*}{\begin{sideways}missing LP\end{sideways}}
&\texttt{clg-msg} & 0.769 & 0.514 & 0.550 & 0.618 & 0.369 & 0.413 \\
&\texttt{email} & 0.769 & 0.460 & 0.380 & 0.713 & 0.107 & 0.185 \\
&\texttt{math-flow} & 0.836 & 0.691 & 0.737 & 0.874 & 0.727 & 0.791 \\
&\texttt{ast-phys} & 0.840 & 0.786 & 0.890 & 0.571 & 0.121 & 0.048 \\
&\texttt{lst-fm} & 0.736 & 0.598 & 0.677 & --- & --- & --- \\
&\texttt{iacontact} & 0.619 & 0.212 & 0.140 & --- & --- & --- \\
&\texttt{forum} & 0.749 & 0.566 & 0.449 & 0.627 & 0.477 & 0.449 \\
&\texttt{topo} & 0.799 & 0.776 & 0.862 & 0.533 & 0.455 & 0.784 \\
&\texttt{act-mooc} & 0.826 & 0.761 & 0.898 & --- & --- & --- \\
&\texttt{arxiv} & 0.685 & 0.552 & 0.692 & --- & --- & --- \\
&\texttt{pow-grid} & 0.563 & -0.009 & -0.080 & 0.622 & 0.335 & 0.358 \\
&\texttt{routers} & 0.812 & 0.133 & 0.131 & 0.717 & 0.243 & 0.197 \\
&\texttt{bio-yeast} & 0.804 & 0.740 & 0.830 & 0.717 & 0.302 & 0.387 \\
&\texttt{fb} & 0.822 & 0.687 & 0.721 & 0.271 & 0.163 & 0.181 \\
&\texttt{blog-cat} & 0.857 & 0.775 & 0.877 & 0.763 & 0.525 & 0.503 \\ \midrule
\multirow{8}{*}{\begin{sideways}future LP\end{sideways}}&
\texttt{clg-msg-t} & 0.815 & 0.601 & 0.691 & 0.817 & 0.669 & 0.669 \\
&\texttt{email-t} & 0.701 & 0.453 & 0.604 & 0.751 & 0.521 & 0.601 \\
&\texttt{math-flow-t} & 0.921 & 0.801 & 0.862 & 0.867 & 0.773 & 0.845 \\
&\texttt{ast-phys-t} & 0.802 & 0.659 & 0.824 & 0.564 & 0.598 & 0.703 \\
&\texttt{lst-fm-t} & 0.798 & 0.625 & 0.677 & 0.642 & 0.312 & 0.319 \\
&\texttt{iacontact-t} & 0.725 & 0.501 & 0.461 & --- & --- & --- \\
&\texttt{forum-t} & 0.785 & 0.576 & 0.568 & 0.830 & 0.685 & 0.708 \\
&\texttt{topo-t} & 0.866 & 0.795 & 0.892 & 0.738 & 0.623 & 0.763 \\
&\texttt{act-mooc-t} & 0.839 & 0.788 & 0.913 & 0.280 & 0.302 & 0.161 \\
\bottomrule
\end{tabular}
\end{adjustbox}
\label{tab:Cls-imb-P/2}
\end{table*}

\subsection{Summary}
Our proposed experimental setup allowed us to conduct the tests presented in this section in order to evaluate the existing LP methods in a controlled environment. Our main findings are: (a) LP methods perform better when future LP is considered than that of missing LP; (b) behavior of the LP methods are not consistent on the two test sets: two-hop and three-hop away node pairs; (c) when absolute performance of the LP methods are compared based on their category, we found significant difference in performance for the test set consisting of two-hop away node pairs, but for its three-hop away counterpart, no significant difference was observed; (d) converting directed networks in undirected ones and applying LP methods on it may significantly affect its performance for future LP on the test set of two-hop away node pairs; (e) injecting imbalance in the test set affects the relative performance of the LP methods for the skew sensitive metrics like AUPR and $Pr@P$, and the rankings based on AUPR and $Pr@P$ are not consistent with that of the skew insensitive metric AUROC; (f) rankings of the LP methods differ based on their performance in the early retrieval phase when imbalanced test set used as opposed to balanced one.

Based on our findings, we recommend evaluating the LP methods in the controlled experimental setup presented in this paper. The methods should be evaluated on both the problem types, on the carefully designed distance controlled and skew controlled test set. A method's performance needs to be compared within its category in this setup, as well as with all the methods irrespective of their categories. We recommend AUPR as a single point summary evaluation measure to be employed on both of the balanced and imbalanced test set. Along with it, PRC, on both of balanced and imbalanced test set, should be investigated to focus on certain portion of the whole spectrum of the thresholds, such as the early retrieval region, which may be necessary for certain applications.
\section{Conclusion and Future Directions}
\label{Concl}
In this paper, we identified a number of gaps in the existing link prediction literature concerning its performance evaluation. This paper proposed an experimental setup which provides a controlled environment based on factors like prediction type, network type, method type, distance between end nodes, class imbalance, evaluation metric, early retrieval scenario, etc., to evaluate LP methods in a rigorous manner. We performed extensive experiments on real network datasets using existing LP methods, and tested carefully designed hypotheses to gain insight on the effect of those factors on LP methods. Subsequently, we presented our recommendations to be followed as best practice toward effective evaluation of LP methods.

This paper considered evaluating link prediction on a simple, unweighted, undirected and homogeneous graph, and analyzed the LP methods which ignore time information. This work can be extended to identify the unexplored avenues in evaluation of LP on specialized graphs, such as weighted, directed and heterogeneous. Various aspects of the performance of temporal LP methods and LP methods, which are specifically designed for the directed networks, may be investigated as further studies.

\section{Declarations}
\subsection{Ethical Approval} This declaration is “not applicable”.
\subsection{Funding} This declaration is “not applicable”.

\bibliography{lp1}


\begin{thebibliography}{65}
\ifx \bisbn   \undefined \def \bisbn  #1{ISBN #1}\fi
\ifx \binits  \undefined \def \binits#1{#1}\fi
\ifx \bauthor  \undefined \def \bauthor#1{#1}\fi
\ifx \batitle  \undefined \def \batitle#1{#1}\fi
\ifx \bjtitle  \undefined \def \bjtitle#1{#1}\fi
\ifx \bvolume  \undefined \def \bvolume#1{\textbf{#1}}\fi
\ifx \byear  \undefined \def \byear#1{#1}\fi
\ifx \bissue  \undefined \def \bissue#1{#1}\fi
\ifx \bfpage  \undefined \def \bfpage#1{#1}\fi
\ifx \blpage  \undefined \def \blpage #1{#1}\fi
\ifx \burl  \undefined \def \burl#1{\textsf{#1}}\fi
\ifx \doiurl  \undefined \def \doiurl#1{\url{https://doi.org/#1}}\fi
\ifx \betal  \undefined \def \betal{\textit{et al.}}\fi
\ifx \binstitute  \undefined \def \binstitute#1{#1}\fi
\ifx \binstitutionaled  \undefined \def \binstitutionaled#1{#1}\fi
\ifx \bctitle  \undefined \def \bctitle#1{#1}\fi
\ifx \beditor  \undefined \def \beditor#1{#1}\fi
\ifx \bpublisher  \undefined \def \bpublisher#1{#1}\fi
\ifx \bbtitle  \undefined \def \bbtitle#1{#1}\fi
\ifx \bedition  \undefined \def \bedition#1{#1}\fi
\ifx \bseriesno  \undefined \def \bseriesno#1{#1}\fi
\ifx \blocation  \undefined \def \blocation#1{#1}\fi
\ifx \bsertitle  \undefined \def \bsertitle#1{#1}\fi
\ifx \bsnm \undefined \def \bsnm#1{#1}\fi
\ifx \bsuffix \undefined \def \bsuffix#1{#1}\fi
\ifx \bparticle \undefined \def \bparticle#1{#1}\fi
\ifx \barticle \undefined \def \barticle#1{#1}\fi
\bibcommenthead
\ifx \bconfdate \undefined \def \bconfdate #1{#1}\fi
\ifx \botherref \undefined \def \botherref #1{#1}\fi
\ifx \url \undefined \def \url#1{\textsf{#1}}\fi
\ifx \bchapter \undefined \def \bchapter#1{#1}\fi
\ifx \bbook \undefined \def \bbook#1{#1}\fi
\ifx \bcomment \undefined \def \bcomment#1{#1}\fi
\ifx \oauthor \undefined \def \oauthor#1{#1}\fi
\ifx \citeauthoryear \undefined \def \citeauthoryear#1{#1}\fi
\ifx \endbibitem  \undefined \def \endbibitem {}\fi
\ifx \bconflocation  \undefined \def \bconflocation#1{#1}\fi
\ifx \arxivurl  \undefined \def \arxivurl#1{\textsf{#1}}\fi
\csname PreBibitemsHook\endcsname

\bibitem[\protect\citeauthoryear{Liben-Nowell and
  Kleinberg}{2007}]{liben2007link}
\begin{barticle}
\bauthor{\bsnm{Liben-Nowell}, \binits{D.}},
\bauthor{\bsnm{Kleinberg}, \binits{J.}}:
\batitle{The link-prediction problem for social networks}.
\bjtitle{Journal of the American society for information science and
  technology}
\bvolume{58}(\bissue{7}),
\bfpage{1019}--\blpage{1031}
(\byear{2007})
\end{barticle}
\endbibitem

\bibitem[\protect\citeauthoryear{Kumar et~al.}{2020}]{kumar2020link}
\begin{barticle}
\bauthor{\bsnm{Kumar}, \binits{A.}},
\bauthor{\bsnm{Singh}, \binits{S.S.}},
\bauthor{\bsnm{Singh}, \binits{K.}},
\bauthor{\bsnm{Biswas}, \binits{B.}}:
\batitle{Link prediction techniques, applications, and performance: A survey}.
\bjtitle{Physica A: Statistical Mechanics and its Applications}
\bvolume{553},
\bfpage{124289}
(\byear{2020})
\end{barticle}
\endbibitem

\bibitem[\protect\citeauthoryear{Lichtenwalter
  et~al.}{2010}]{lichtenwalter2010new}
\begin{bchapter}
\bauthor{\bsnm{Lichtenwalter}, \binits{R.N.}},
\bauthor{\bsnm{Lussier}, \binits{J.T.}},
\bauthor{\bsnm{Chawla}, \binits{N.V.}}:
\bctitle{New perspectives and methods in link prediction}.
In: \bbtitle{Proceedings of the 16th ACM SIGKDD International Conference on
  Knowledge Discovery and Data Mining},
pp. \bfpage{243}--\blpage{252}
(\byear{2010})
\end{bchapter}
\endbibitem

\bibitem[\protect\citeauthoryear{Daud et~al.}{2020}]{daud2020applications}
\begin{barticle}
\bauthor{\bsnm{Daud}, \binits{N.N.}},
\bauthor{\bsnm{Ab~Hamid}, \binits{S.H.}},
\bauthor{\bsnm{Saadoon}, \binits{M.}},
\bauthor{\bsnm{Sahran}, \binits{F.}},
\bauthor{\bsnm{Anuar}, \binits{N.B.}}:
\batitle{Applications of link prediction in social networks: A review}.
\bjtitle{Journal of Network and Computer Applications}
\bvolume{166},
\bfpage{102716}
(\byear{2020})
\end{barticle}
\endbibitem

\bibitem[\protect\citeauthoryear{Calderoni et~al.}{2020}]{calderoni2020robust}
\begin{barticle}
\bauthor{\bsnm{Calderoni}, \binits{F.}},
\bauthor{\bsnm{Catanese}, \binits{S.}},
\bauthor{\bsnm{De~Meo}, \binits{P.}},
\bauthor{\bsnm{Ficara}, \binits{A.}},
\bauthor{\bsnm{Fiumara}, \binits{G.}}:
\batitle{Robust link prediction in criminal networks: A case study of the
  sicilian mafia}.
\bjtitle{Expert Systems with Applications}
\bvolume{161},
\bfpage{113666}
(\byear{2020})
\end{barticle}
\endbibitem

\bibitem[\protect\citeauthoryear{Berlusconi et~al.}{2016}]{berlusconi2016link}
\begin{barticle}
\bauthor{\bsnm{Berlusconi}, \binits{G.}},
\bauthor{\bsnm{Calderoni}, \binits{F.}},
\bauthor{\bsnm{Parolini}, \binits{N.}},
\bauthor{\bsnm{Verani}, \binits{M.}},
\bauthor{\bsnm{Piccardi}, \binits{C.}}:
\batitle{Link prediction in criminal networks: A tool for criminal intelligence
  analysis}.
\bjtitle{PloS one}
\bvolume{11}(\bissue{4}),
\bfpage{0154244}
(\byear{2016})
\end{barticle}
\endbibitem

\bibitem[\protect\citeauthoryear{Sett et~al.}{2018}]{sett2018temporal}
\begin{barticle}
\bauthor{\bsnm{Sett}, \binits{N.}},
\bauthor{\bsnm{Basu}, \binits{S.}},
\bauthor{\bsnm{Nandi}, \binits{S.}},
\bauthor{\bsnm{Singh}, \binits{S.R.}}:
\batitle{Temporal link prediction in multi-relational network}.
\bjtitle{World Wide Web}
\bvolume{21},
\bfpage{395}--\blpage{419}
(\byear{2018})
\end{barticle}
\endbibitem

\bibitem[\protect\citeauthoryear{Sun et~al.}{2011}]{sun2011co}
\begin{bchapter}
\bauthor{\bsnm{Sun}, \binits{Y.}},
\bauthor{\bsnm{Barber}, \binits{R.}},
\bauthor{\bsnm{Gupta}, \binits{M.}},
\bauthor{\bsnm{Aggarwal}, \binits{C.C.}},
\bauthor{\bsnm{Han}, \binits{J.}}:
\bctitle{Co-author relationship prediction in heterogeneous bibliographic
  networks}.
In: \bbtitle{2011 International Conference on Advances in Social Networks
  Analysis and Mining},
pp. \bfpage{121}--\blpage{128}
(\byear{2011}).
\bcomment{IEEE}
\end{bchapter}
\endbibitem

\bibitem[\protect\citeauthoryear{Chuan et~al.}{2018}]{chuan2018link}
\begin{barticle}
\bauthor{\bsnm{Chuan}, \binits{P.M.}},
\bauthor{\bsnm{Son}, \binits{L.H.}},
\bauthor{\bsnm{Ali}, \binits{M.}},
\bauthor{\bsnm{Khang}, \binits{T.D.}},
\bauthor{\bsnm{Huong}, \binits{L.T.}},
\bauthor{\bsnm{Dey}, \binits{N.}}:
\batitle{Link prediction in co-authorship networks based on hybrid content
  similarity metric}.
\bjtitle{Applied Intelligence}
\bvolume{48},
\bfpage{2470}--\blpage{2486}
(\byear{2018})
\end{barticle}
\endbibitem

\bibitem[\protect\citeauthoryear{Huang et~al.}{2005}]{huang2005link}
\begin{bchapter}
\bauthor{\bsnm{Huang}, \binits{Z.}},
\bauthor{\bsnm{Li}, \binits{X.}},
\bauthor{\bsnm{Chen}, \binits{H.}}:
\bctitle{Link prediction approach to collaborative filtering}.
In: \bbtitle{Proceedings of the 5th ACM/IEEE-CS Joint Conference on Digital
  Libraries},
pp. \bfpage{141}--\blpage{142}
(\byear{2005})
\end{bchapter}
\endbibitem

\bibitem[\protect\citeauthoryear{Yilmaz et~al.}{2023}]{yilmaz2023link}
\begin{barticle}
\bauthor{\bsnm{Yilmaz}, \binits{E.A.}},
\bauthor{\bsnm{Balcisoy}, \binits{S.}},
\bauthor{\bsnm{Bozkaya}, \binits{B.}}:
\batitle{A link prediction-based recommendation system using transactional
  data}.
\bjtitle{Scientific Reports}
\bvolume{13}(\bissue{1}),
\bfpage{6905}
(\byear{2023})
\end{barticle}
\endbibitem

\bibitem[\protect\citeauthoryear{Lee and Zhou}{2021}]{lee2021collaborative}
\begin{barticle}
\bauthor{\bsnm{Lee}, \binits{Y.-L.}},
\bauthor{\bsnm{Zhou}, \binits{T.}}:
\batitle{Collaborative filtering approach to link prediction}.
\bjtitle{Physica A: Statistical Mechanics and its Applications}
\bvolume{578},
\bfpage{126107}
(\byear{2021})
\end{barticle}
\endbibitem

\bibitem[\protect\citeauthoryear{Wang et~al.}{2014}]{wang2014link}
\begin{botherref}
\oauthor{\bsnm{Wang}, \binits{P.}},
\oauthor{\bsnm{Xu}, \binits{B.}},
\oauthor{\bsnm{Wu}, \binits{Y.}},
\oauthor{\bsnm{Zhou}, \binits{X.}}:
Link prediction in social networks: the state-of-the-art.
arXiv preprint arXiv:1411.5118
(2014)
\end{botherref}
\endbibitem

\bibitem[\protect\citeauthoryear{Vahidi~Farashah
  et~al.}{2021}]{vahidi2021hybrid}
\begin{barticle}
\bauthor{\bsnm{Vahidi~Farashah}, \binits{M.}},
\bauthor{\bsnm{Etebarian}, \binits{A.}},
\bauthor{\bsnm{Azmi}, \binits{R.}},
\bauthor{\bsnm{Ebrahimzadeh~Dastjerdi}, \binits{R.}}:
\batitle{A hybrid recommender system based-on link prediction for movie baskets
  analysis}.
\bjtitle{Journal of Big Data}
\bvolume{8},
\bfpage{1}--\blpage{24}
(\byear{2021})
\end{barticle}
\endbibitem

\bibitem[\protect\citeauthoryear{Leroy et~al.}{2010}]{leroy2010cold}
\begin{bchapter}
\bauthor{\bsnm{Leroy}, \binits{V.}},
\bauthor{\bsnm{Cambazoglu}, \binits{B.B.}},
\bauthor{\bsnm{Bonchi}, \binits{F.}}:
\bctitle{Cold start link prediction}.
In: \bbtitle{Proceedings of the 16th ACM SIGKDD International Conference on
  Knowledge Discovery and Data Mining},
pp. \bfpage{393}--\blpage{402}
(\byear{2010})
\end{bchapter}
\endbibitem

\bibitem[\protect\citeauthoryear{Li et~al.}{2023}]{li2023effective}
\begin{barticle}
\bauthor{\bsnm{Li}, \binits{W.}},
\bauthor{\bsnm{Li}, \binits{T.}},
\bauthor{\bsnm{Berahmand}, \binits{K.}}:
\batitle{An effective link prediction method in multiplex social networks using
  local random walk towards dependable pathways}.
\bjtitle{Journal of Combinatorial Optimization}
\bvolume{45}(\bissue{1}),
\bfpage{31}
(\byear{2023})
\end{barticle}
\endbibitem

\bibitem[\protect\citeauthoryear{Zhou et~al.}{2009}]{zhou2009predicting}
\begin{barticle}
\bauthor{\bsnm{Zhou}, \binits{T.}},
\bauthor{\bsnm{L{\"u}}, \binits{L.}},
\bauthor{\bsnm{Zhang}, \binits{Y.-C.}}:
\batitle{Predicting missing links via local information}.
\bjtitle{The European Physical Journal B}
\bvolume{71},
\bfpage{623}--\blpage{630}
(\byear{2009})
\end{barticle}
\endbibitem

\bibitem[\protect\citeauthoryear{Zhang and Chen}{2018}]{zhang2018link}
\begin{botherref}
\oauthor{\bsnm{Zhang}, \binits{M.}},
\oauthor{\bsnm{Chen}, \binits{Y.}}:
Link prediction based on graph neural networks.
Advances in neural information processing systems
\textbf{31}
(2018)
\end{botherref}
\endbibitem

\bibitem[\protect\citeauthoryear{Perozzi et~al.}{2014}]{perozzi2014deepwalk}
\begin{bchapter}
\bauthor{\bsnm{Perozzi}, \binits{B.}},
\bauthor{\bsnm{Al-Rfou}, \binits{R.}},
\bauthor{\bsnm{Skiena}, \binits{S.}}:
\bctitle{Deepwalk: Online learning of social representations}.
In: \bbtitle{Proceedings of the 20th ACM SIGKDD International Conference on
  Knowledge Discovery and Data Mining},
pp. \bfpage{701}--\blpage{710}
(\byear{2014})
\end{bchapter}
\endbibitem

\bibitem[\protect\citeauthoryear{Grover and
  Leskovec}{2016}]{grover2016node2vec}
\begin{bchapter}
\bauthor{\bsnm{Grover}, \binits{A.}},
\bauthor{\bsnm{Leskovec}, \binits{J.}}:
\bctitle{node2vec: Scalable feature learning for networks}.
In: \bbtitle{Proceedings of the 22nd ACM SIGKDD International Conference on
  Knowledge Discovery and Data Mining},
pp. \bfpage{855}--\blpage{864}
(\byear{2016})
\end{bchapter}
\endbibitem

\bibitem[\protect\citeauthoryear{Hamilton et~al.}{2017}]{hamilton2017inductive}
\begin{botherref}
\oauthor{\bsnm{Hamilton}, \binits{W.}},
\oauthor{\bsnm{Ying}, \binits{Z.}},
\oauthor{\bsnm{Leskovec}, \binits{J.}}:
Inductive representation learning on large graphs.
Advances in neural information processing systems
\textbf{30}
(2017)
\end{botherref}
\endbibitem

\bibitem[\protect\citeauthoryear{He et~al.}{2020}]{he2020lightgcn}
\begin{bchapter}
\bauthor{\bsnm{He}, \binits{X.}},
\bauthor{\bsnm{Deng}, \binits{K.}},
\bauthor{\bsnm{Wang}, \binits{X.}},
\bauthor{\bsnm{Li}, \binits{Y.}},
\bauthor{\bsnm{Zhang}, \binits{Y.}},
\bauthor{\bsnm{Wang}, \binits{M.}}:
\bctitle{Lightgcn: Simplifying and powering graph convolution network for
  recommendation}.
In: \bbtitle{Proceedings of the 43rd International ACM SIGIR Conference on
  Research and Development in Information Retrieval},
pp. \bfpage{639}--\blpage{648}
(\byear{2020})
\end{bchapter}
\endbibitem

\bibitem[\protect\citeauthoryear{Menon and Elkan}{2011}]{menon2011link}
\begin{bchapter}
\bauthor{\bsnm{Menon}, \binits{A.K.}},
\bauthor{\bsnm{Elkan}, \binits{C.}}:
\bctitle{Link prediction via matrix factorization}.
In: \bbtitle{Machine Learning and Knowledge Discovery in Databases: European
  Conference, ECML PKDD 2011, Athens, Greece, September 5-9, 2011, Proceedings,
  Part II 22},
pp. \bfpage{437}--\blpage{452}
(\byear{2011}).
\bcomment{Springer}
\end{bchapter}
\endbibitem

\bibitem[\protect\citeauthoryear{Mart{\'\i}nez
  et~al.}{2016}]{martinez2016survey}
\begin{barticle}
\bauthor{\bsnm{Mart{\'\i}nez}, \binits{V.}},
\bauthor{\bsnm{Berzal}, \binits{F.}},
\bauthor{\bsnm{Cubero}, \binits{J.-C.}}:
\batitle{A survey of link prediction in complex networks}.
\bjtitle{ACM computing surveys (CSUR)}
\bvolume{49}(\bissue{4}),
\bfpage{1}--\blpage{33}
(\byear{2016})
\end{barticle}
\endbibitem

\bibitem[\protect\citeauthoryear{Qin and Yeung}{2023}]{qin2023temporal}
\begin{barticle}
\bauthor{\bsnm{Qin}, \binits{M.}},
\bauthor{\bsnm{Yeung}, \binits{D.-Y.}}:
\batitle{Temporal link prediction: A unified framework, taxonomy, and review}.
\bjtitle{ACM Computing Surveys}
\bvolume{56}(\bissue{4}),
\bfpage{1}--\blpage{40}
(\byear{2023})
\end{barticle}
\endbibitem

\bibitem[\protect\citeauthoryear{Granovetter}{1973}]{granovetter1973strength}
\begin{barticle}
\bauthor{\bsnm{Granovetter}, \binits{M.S.}}:
\batitle{The strength of weak ties}.
\bjtitle{American journal of sociology}
\bvolume{78}(\bissue{6}),
\bfpage{1360}--\blpage{1380}
(\byear{1973})
\end{barticle}
\endbibitem

\bibitem[\protect\citeauthoryear{Cai and Ji}{2020}]{cai2020multi}
\begin{bchapter}
\bauthor{\bsnm{Cai}, \binits{L.}},
\bauthor{\bsnm{Ji}, \binits{S.}}:
\bctitle{A multi-scale approach for graph link prediction}.
In: \bbtitle{Proceedings of the AAAI Conference on Artificial Intelligence},
vol. \bseriesno{34},
pp. \bfpage{3308}--\blpage{3315}
(\byear{2020})
\end{bchapter}
\endbibitem

\bibitem[\protect\citeauthoryear{Pan et~al.}{2021}]{pan2021neural}
\begin{botherref}
\oauthor{\bsnm{Pan}, \binits{L.}},
\oauthor{\bsnm{Shi}, \binits{C.}},
\oauthor{\bsnm{Dokmani{\'c}}, \binits{I.}}:
Neural link prediction with walk pooling.
arXiv preprint arXiv:2110.04375
(2021)
\end{botherref}
\endbibitem

\bibitem[\protect\citeauthoryear{Tan et~al.}{2023}]{tan2023bring}
\begin{bchapter}
\bauthor{\bsnm{Tan}, \binits{Q.}},
\bauthor{\bsnm{Zhang}, \binits{X.}},
\bauthor{\bsnm{Liu}, \binits{N.}},
\bauthor{\bsnm{Zha}, \binits{D.}},
\bauthor{\bsnm{Li}, \binits{L.}},
\bauthor{\bsnm{Chen}, \binits{R.}},
\bauthor{\bsnm{Choi}, \binits{S.-H.}},
\bauthor{\bsnm{Hu}, \binits{X.}}:
\bctitle{Bring your own view: Graph neural networks for link prediction with
  personalized subgraph selection}.
In: \bbtitle{Proceedings of the Sixteenth ACM International Conference on Web
  Search and Data Mining},
pp. \bfpage{625}--\blpage{633}
(\byear{2023})
\end{bchapter}
\endbibitem

\bibitem[\protect\citeauthoryear{Strogatz}{2001}]{strogatz2001exploring}
\begin{barticle}
\bauthor{\bsnm{Strogatz}, \binits{S.H.}}:
\batitle{Exploring complex networks}.
\bjtitle{nature}
\bvolume{410}(\bissue{6825}),
\bfpage{268}--\blpage{276}
(\byear{2001})
\end{barticle}
\endbibitem

\bibitem[\protect\citeauthoryear{Hanley and McNeil}{1982}]{hanley1982meaning}
\begin{barticle}
\bauthor{\bsnm{Hanley}, \binits{J.A.}},
\bauthor{\bsnm{McNeil}, \binits{B.J.}}:
\batitle{The meaning and use of the area under a receiver operating
  characteristic (roc) curve.}
\bjtitle{Radiology}
\bvolume{143}(\bissue{1}),
\bfpage{29}--\blpage{36}
(\byear{1982})
\end{barticle}
\endbibitem

\bibitem[\protect\citeauthoryear{Saito and
  Rehmsmeier}{2015}]{saito2015precision}
\begin{barticle}
\bauthor{\bsnm{Saito}, \binits{T.}},
\bauthor{\bsnm{Rehmsmeier}, \binits{M.}}:
\batitle{The precision-recall plot is more informative than the roc plot when
  evaluating binary classifiers on imbalanced datasets}.
\bjtitle{PloS one}
\bvolume{10}(\bissue{3}),
\bfpage{0118432}
(\byear{2015})
\end{barticle}
\endbibitem

\bibitem[\protect\citeauthoryear{Truchon and
  Bayly}{2007}]{truchon2007evaluating}
\begin{barticle}
\bauthor{\bsnm{Truchon}, \binits{J.-F.}},
\bauthor{\bsnm{Bayly}, \binits{C.I.}}:
\batitle{Evaluating virtual screening methods: good and bad metrics for the
  “early recognition” problem}.
\bjtitle{Journal of chemical information and modeling}
\bvolume{47}(\bissue{2}),
\bfpage{488}--\blpage{508}
(\byear{2007})
\end{barticle}
\endbibitem

\bibitem[\protect\citeauthoryear{Swamidass et~al.}{2010}]{swamidass2010croc}
\begin{barticle}
\bauthor{\bsnm{Swamidass}, \binits{S.J.}},
\bauthor{\bsnm{Azencott}, \binits{C.-A.}},
\bauthor{\bsnm{Daily}, \binits{K.}},
\bauthor{\bsnm{Baldi}, \binits{P.}}:
\batitle{A croc stronger than roc: measuring, visualizing and optimizing early
  retrieval}.
\bjtitle{Bioinformatics}
\bvolume{26}(\bissue{10}),
\bfpage{1348}--\blpage{1356}
(\byear{2010})
\end{barticle}
\endbibitem

\bibitem[\protect\citeauthoryear{Yang et~al.}{2015}]{yang2015evaluating}
\begin{barticle}
\bauthor{\bsnm{Yang}, \binits{Y.}},
\bauthor{\bsnm{Lichtenwalter}, \binits{R.N.}},
\bauthor{\bsnm{Chawla}, \binits{N.V.}}:
\batitle{Evaluating link prediction methods}.
\bjtitle{Knowledge and Information Systems}
\bvolume{45},
\bfpage{751}--\blpage{782}
(\byear{2015})
\end{barticle}
\endbibitem

\bibitem[\protect\citeauthoryear{Chen et~al.}{2017}]{chen2017link}
\begin{barticle}
\bauthor{\bsnm{Chen}, \binits{B.}},
\bauthor{\bsnm{Li}, \binits{F.}},
\bauthor{\bsnm{Chen}, \binits{S.}},
\bauthor{\bsnm{Hu}, \binits{R.}},
\bauthor{\bsnm{Chen}, \binits{L.}}:
\batitle{Link prediction based on non-negative matrix factorization}.
\bjtitle{PloS one}
\bvolume{12}(\bissue{8}),
\bfpage{0182968}
(\byear{2017})
\end{barticle}
\endbibitem

\bibitem[\protect\citeauthoryear{Kunegis et~al.}{2010}]{kunegis2010link}
\begin{bchapter}
\bauthor{\bsnm{Kunegis}, \binits{J.}},
\bauthor{\bsnm{De~Luca}, \binits{E.W.}},
\bauthor{\bsnm{Albayrak}, \binits{S.}}:
\bctitle{The link prediction problem in bipartite networks}.
In: \bbtitle{International Conference on Information Processing and Management
  of Uncertainty in Knowledge-based Systems},
pp. \bfpage{380}--\blpage{389}
(\byear{2010}).
\bcomment{Springer}
\end{bchapter}
\endbibitem

\bibitem[\protect\citeauthoryear{{\"O}zer et~al.}{2024}]{ozer2024link}
\begin{bchapter}
\bauthor{\bsnm{{\"O}zer}, \binits{{\c{S}}.D.I.}},
\bauthor{\bsnm{Orman}, \binits{G.K.}},
\bauthor{\bsnm{Labatut}, \binits{V.}}:
\bctitle{Link prediction in bipartite networks}.
In: \bbtitle{28th International Conference on Knowledge-Based and Intelligent
  Information \& Engineering Systems (KES)}
(\byear{2024})
\end{bchapter}
\endbibitem

\bibitem[\protect\citeauthoryear{Li et~al.}{2018}]{li2018link}
\begin{barticle}
\bauthor{\bsnm{Li}, \binits{J.-c.}},
\bauthor{\bsnm{Zhao}, \binits{D.-l.}},
\bauthor{\bsnm{Ge}, \binits{B.-F.}},
\bauthor{\bsnm{Yang}, \binits{K.-W.}},
\bauthor{\bsnm{Chen}, \binits{Y.-W.}}:
\batitle{A link prediction method for heterogeneous networks based on bp neural
  network}.
\bjtitle{Physica A: Statistical Mechanics and its Applications}
\bvolume{495},
\bfpage{1}--\blpage{17}
(\byear{2018})
\end{barticle}
\endbibitem

\bibitem[\protect\citeauthoryear{Wang et~al.}{2023}]{wang2023multi}
\begin{barticle}
\bauthor{\bsnm{Wang}, \binits{H.}},
\bauthor{\bsnm{Cui}, \binits{Z.}},
\bauthor{\bsnm{Liu}, \binits{R.}},
\bauthor{\bsnm{Fang}, \binits{L.}},
\bauthor{\bsnm{Sha}, \binits{Y.}}:
\batitle{A multi-type transferable method for missing link prediction in
  heterogeneous social networks}.
\bjtitle{IEEE Transactions on Knowledge and Data Engineering}
\bvolume{35}(\bissue{11}),
\bfpage{10981}--\blpage{10991}
(\byear{2023})
\end{barticle}
\endbibitem

\bibitem[\protect\citeauthoryear{Schall}{2014}]{schall2014link}
\begin{barticle}
\bauthor{\bsnm{Schall}, \binits{D.}}:
\batitle{Link prediction in directed social networks}.
\bjtitle{Social Network Analysis and Mining}
\bvolume{4}(\bissue{1}),
\bfpage{157}
(\byear{2014})
\end{barticle}
\endbibitem

\bibitem[\protect\citeauthoryear{Sett et~al.}{2018}]{sett2018exploiting}
\begin{barticle}
\bauthor{\bsnm{Sett}, \binits{N.}},
\bauthor{\bsnm{Devesh}},
\bauthor{\bsnm{Singh}, \binits{S.R.}},
\bauthor{\bsnm{Nandi}, \binits{S.}}:
\batitle{Exploiting reciprocity toward link prediction}.
\bjtitle{Knowledge and Information Systems}
\bvolume{55},
\bfpage{1}--\blpage{13}
(\byear{2018})
\end{barticle}
\endbibitem

\bibitem[\protect\citeauthoryear{Lichtnwalter and
  Chawla}{2012}]{lichtnwalter2012link}
\begin{bchapter}
\bauthor{\bsnm{Lichtnwalter}, \binits{R.}},
\bauthor{\bsnm{Chawla}, \binits{N.V.}}:
\bctitle{Link prediction: fair and effective evaluation}.
In: \bbtitle{2012 IEEE/ACM International Conference on Advances in Social
  Networks Analysis and Mining},
pp. \bfpage{376}--\blpage{383}
(\byear{2012}).
\bcomment{IEEE}
\end{bchapter}
\endbibitem

\bibitem[\protect\citeauthoryear{He et~al.}{2024}]{he2024link}
\begin{barticle}
\bauthor{\bsnm{He}, \binits{X.}},
\bauthor{\bsnm{Ghasemian}, \binits{A.}},
\bauthor{\bsnm{Lee}, \binits{E.}},
\bauthor{\bsnm{Schwarze}, \binits{A.C.}},
\bauthor{\bsnm{Clauset}, \binits{A.}},
\bauthor{\bsnm{Mucha}, \binits{P.J.}}:
\batitle{Link prediction accuracy on real-world networks under non-uniform
  missing-edge patterns}.
\bjtitle{Plos one}
\bvolume{19}(\bissue{7}),
\bfpage{0306883}
(\byear{2024})
\end{barticle}
\endbibitem

\bibitem[\protect\citeauthoryear{Junuthula
  et~al.}{2016}]{junuthula2016evaluating}
\begin{bchapter}
\bauthor{\bsnm{Junuthula}, \binits{R.R.}},
\bauthor{\bsnm{Xu}, \binits{K.S.}},
\bauthor{\bsnm{Devabhaktuni}, \binits{V.K.}}:
\bctitle{Evaluating link prediction accuracy in dynamic networks with added and
  removed edges}.
In: \bbtitle{2016 IEEE International Conferences on Big Data and Cloud
  Computing (BDCloud), Social Computing and Networking (SocialCom), Sustainable
  Computing and Communications (SustainCom)(BDCloud-SocialCom-SustainCom)},
pp. \bfpage{377}--\blpage{384}
(\byear{2016}).
\bcomment{IEEE}
\end{bchapter}
\endbibitem

\bibitem[\protect\citeauthoryear{Poursafaei
  et~al.}{2022}]{poursafaei2022towards}
\begin{barticle}
\bauthor{\bsnm{Poursafaei}, \binits{F.}},
\bauthor{\bsnm{Huang}, \binits{S.}},
\bauthor{\bsnm{Pelrine}, \binits{K.}},
\bauthor{\bsnm{Rabbany}, \binits{R.}}:
\batitle{Towards better evaluation for dynamic link prediction}.
\bjtitle{Advances in Neural Information Processing Systems}
\bvolume{35},
\bfpage{32928}--\blpage{32941}
(\byear{2022})
\end{barticle}
\endbibitem

\bibitem[\protect\citeauthoryear{Masrour et~al.}{2020}]{masrour2020bursting}
\begin{bchapter}
\bauthor{\bsnm{Masrour}, \binits{F.}},
\bauthor{\bsnm{Wilson}, \binits{T.}},
\bauthor{\bsnm{Yan}, \binits{H.}},
\bauthor{\bsnm{Tan}, \binits{P.-N.}},
\bauthor{\bsnm{Esfahanian}, \binits{A.}}:
\bctitle{Bursting the filter bubble: Fairness-aware network link prediction}.
In: \bbtitle{Proceedings of the AAAI Conference on Artificial Intelligence},
vol. \bseriesno{34},
pp. \bfpage{841}--\blpage{848}
(\byear{2020})
\end{bchapter}
\endbibitem

\bibitem[\protect\citeauthoryear{Nasiri et~al.}{2023}]{nasiri2023robust}
\begin{barticle}
\bauthor{\bsnm{Nasiri}, \binits{E.}},
\bauthor{\bsnm{Berahmand}, \binits{K.}},
\bauthor{\bsnm{Li}, \binits{Y.}}:
\batitle{Robust graph regularization nonnegative matrix factorization for link
  prediction in attributed networks}.
\bjtitle{Multimedia Tools and Applications}
\bvolume{82}(\bissue{3}),
\bfpage{3745}--\blpage{3768}
(\byear{2023})
\end{barticle}
\endbibitem

\bibitem[\protect\citeauthoryear{Dillon}{1983}]{dillon1983introduction}
\begin{barticle}
\bauthor{\bsnm{Dillon}, \binits{M.}}:
\batitle{Introduction to modern information retrieval}.
\bjtitle{Information Processing \& Management}
\bvolume{19}(\bissue{6}),
\bfpage{402}--\blpage{403}
(\byear{1983})
\end{barticle}
\endbibitem

\bibitem[\protect\citeauthoryear{Adamic and Adar}{2003}]{admic2001friends}
\begin{barticle}
\bauthor{\bsnm{Adamic}, \binits{L.A.}},
\bauthor{\bsnm{Adar}, \binits{E.}}:
\batitle{Friends and neighbors on the web}.
\bjtitle{Social networks}
\bvolume{25}(\bissue{3}),
\bfpage{211}--\blpage{230}
(\byear{2003})
\end{barticle}
\endbibitem

\bibitem[\protect\citeauthoryear{Barab{\^a}si
  et~al.}{2002}]{barabasi2002evolution}
\begin{barticle}
\bauthor{\bsnm{Barab{\^a}si}, \binits{A.-L.}},
\bauthor{\bsnm{Jeong}, \binits{H.}},
\bauthor{\bsnm{N{\'e}da}, \binits{Z.}},
\bauthor{\bsnm{Ravasz}, \binits{E.}},
\bauthor{\bsnm{Schubert}, \binits{A.}},
\bauthor{\bsnm{Vicsek}, \binits{T.}}:
\batitle{Evolution of the social network of scientific collaborations}.
\bjtitle{Physica A: Statistical mechanics and its applications}
\bvolume{311}(\bissue{3-4}),
\bfpage{590}--\blpage{614}
(\byear{2002})
\end{barticle}
\endbibitem

\bibitem[\protect\citeauthoryear{Newman}{2001}]{newman2001clustering}
\begin{barticle}
\bauthor{\bsnm{Newman}, \binits{M.E.}}:
\batitle{Clustering and preferential attachment in growing networks}.
\bjtitle{Physical review E}
\bvolume{64}(\bissue{2}),
\bfpage{025102}
(\byear{2001})
\end{barticle}
\endbibitem

\bibitem[\protect\citeauthoryear{Katz}{1953}]{katz1953new}
\begin{barticle}
\bauthor{\bsnm{Katz}, \binits{L.}}:
\batitle{A new status index derived from sociometric analysis}.
\bjtitle{Psychometrika}
\bvolume{18}(\bissue{1}),
\bfpage{39}--\blpage{43}
(\byear{1953})
\end{barticle}
\endbibitem

\bibitem[\protect\citeauthoryear{Mikolov et~al.}{2013}]{mikolov2013distributed}
\begin{botherref}
\oauthor{\bsnm{Mikolov}, \binits{T.}},
\oauthor{\bsnm{Sutskever}, \binits{I.}},
\oauthor{\bsnm{Chen}, \binits{K.}},
\oauthor{\bsnm{Corrado}, \binits{G.S.}},
\oauthor{\bsnm{Dean}, \binits{J.}}:
Distributed representations of words and phrases and their compositionality.
Advances in neural information processing systems
\textbf{26}
(2013)
\end{botherref}
\endbibitem

\bibitem[\protect\citeauthoryear{Morin and
  Bengio}{2005}]{morin2005hierarchical}
\begin{bchapter}
\bauthor{\bsnm{Morin}, \binits{F.}},
\bauthor{\bsnm{Bengio}, \binits{Y.}}:
\bctitle{Hierarchical probabilistic neural network language model}.
In: \bbtitle{International Workshop on Artificial Intelligence and Statistics},
pp. \bfpage{246}--\blpage{252}
(\byear{2005}).
\bcomment{PMLR}
\end{bchapter}
\endbibitem

\bibitem[\protect\citeauthoryear{Mnih and Hinton}{2008}]{mnih2008scalable}
\begin{botherref}
\oauthor{\bsnm{Mnih}, \binits{A.}},
\oauthor{\bsnm{Hinton}, \binits{G.E.}}:
A scalable hierarchical distributed language model.
Advances in neural information processing systems
\textbf{21}
(2008)
\end{botherref}
\endbibitem

\bibitem[\protect\citeauthoryear{Panzarasa
  et~al.}{2009}]{panzarasa2009patterns}
\begin{barticle}
\bauthor{\bsnm{Panzarasa}, \binits{P.}},
\bauthor{\bsnm{Opsahl}, \binits{T.}},
\bauthor{\bsnm{Carley}, \binits{K.M.}}:
\batitle{Patterns and dynamics of users' behavior and interaction: Network
  analysis of an online community}.
\bjtitle{Journal of the American Society for Information Science and
  Technology}
\bvolume{60}(\bissue{5}),
\bfpage{911}--\blpage{932}
(\byear{2009})
\end{barticle}
\endbibitem

\bibitem[\protect\citeauthoryear{Paranjape et~al.}{2017}]{paranjape2017motifs}
\begin{bchapter}
\bauthor{\bsnm{Paranjape}, \binits{A.}},
\bauthor{\bsnm{Benson}, \binits{A.R.}},
\bauthor{\bsnm{Leskovec}, \binits{J.}}:
\bctitle{Motifs in temporal networks}.
In: \bbtitle{Proceedings of the Tenth ACM International Conference on Web
  Search and Data Mining},
pp. \bfpage{601}--\blpage{610}
(\byear{2017})
\end{bchapter}
\endbibitem

\bibitem[\protect\citeauthoryear{Leskovec et~al.}{2007}]{leskovec2007graph}
\begin{barticle}
\bauthor{\bsnm{Leskovec}, \binits{J.}},
\bauthor{\bsnm{Kleinberg}, \binits{J.}},
\bauthor{\bsnm{Faloutsos}, \binits{C.}}:
\batitle{Graph evolution: Densification and shrinking diameters}.
\bjtitle{ACM transactions on Knowledge Discovery from Data (TKDD)}
\bvolume{1}(\bissue{1}),
\bfpage{2}
(\byear{2007})
\end{barticle}
\endbibitem

\bibitem[\protect\citeauthoryear{Kumar et~al.}{2019}]{kumar2019predicting}
\begin{bchapter}
\bauthor{\bsnm{Kumar}, \binits{S.}},
\bauthor{\bsnm{Zhang}, \binits{X.}},
\bauthor{\bsnm{Leskovec}, \binits{J.}}:
\bctitle{Predicting dynamic embedding trajectory in temporal interaction
  networks}.
In: \bbtitle{Proceedings of the 25th ACM SIGKDD International Conference on
  Knowledge Discovery \& Data Mining},
pp. \bfpage{1269}--\blpage{1278}
(\byear{2019})
\end{bchapter}
\endbibitem

\bibitem[\protect\citeauthoryear{Rossi and Ahmed}{2015}]{nr}
\begin{bchapter}
\bauthor{\bsnm{Rossi}, \binits{R.A.}},
\bauthor{\bsnm{Ahmed}, \binits{N.K.}}:
\bctitle{The network data repository with interactive graph analytics and
  visualization}.
In: \bbtitle{AAAI}
(\byear{2015}).
\burl{https://networkrepository.com}
\end{bchapter}
\endbibitem

\bibitem[\protect\citeauthoryear{Kunegis}{2013}]{konect}
\begin{bchapter}
\bauthor{\bsnm{Kunegis}, \binits{J.}}:
\bctitle{{KONECT} -- {The} {Koblenz} {Network} {Collection}}.
In: \bbtitle{Proc. Int. Conf. on World Wide Web Companion},
pp. \bfpage{1343}--\blpage{1350}
(\byear{2013}).
\burl{http://dl.acm.org/citation.cfm?id=2488173}
\end{bchapter}
\endbibitem

\bibitem[\protect\citeauthoryear{Zhang et~al.}{2005}]{konect:zhang05}
\begin{barticle}
\bauthor{\bsnm{Zhang}, \binits{B.}},
\bauthor{\bsnm{Liu}, \binits{R.}},
\bauthor{\bsnm{Massey}, \binits{D.}},
\bauthor{\bsnm{Zhang}, \binits{L.}}:
\batitle{Collecting the {Internet} {AS}-level topology}.
\bjtitle{SIGCOMM Comput. Communication Review}
\bvolume{35}(\bissue{1}),
\bfpage{53}--\blpage{61}
(\byear{2005})
\end{barticle}
\endbibitem

\bibitem[\protect\citeauthoryear{Leskovec and
  Mcauley}{2012}]{leskovec2012learning}
\begin{botherref}
\oauthor{\bsnm{Leskovec}, \binits{J.}},
\oauthor{\bsnm{Mcauley}, \binits{J.}}:
Learning to discover social circles in ego networks.
Advances in neural information processing systems
\textbf{25}
(2012)
\end{botherref}
\endbibitem

\bibitem[\protect\citeauthoryear{Kerrache et~al.}{2020}]{kerrache2020scalable}
\begin{barticle}
\bauthor{\bsnm{Kerrache}, \binits{S.}},
\bauthor{\bsnm{Alharbi}, \binits{R.}},
\bauthor{\bsnm{Benhidour}, \binits{H.}}:
\batitle{A scalable similarity-popularity link prediction method}.
\bjtitle{Scientific reports}
\bvolume{10}(\bissue{1}),
\bfpage{6394}
(\byear{2020})
\end{barticle}
\endbibitem

\end{thebibliography}
 \begin{appendices}

\section{Consolidated results: Best performing methods in individual groups of methods}\label{Ap-A}

\begin{table*}[h]
\centering
\caption{Best performing methods in each group: local-sim, global-sim and learning, for future LP in test set of two-hop away node pairs}
\renewcommand{\arraystretch}{1.7} 
\begin{adjustbox}{width=0.9\textwidth,center}
\begin{tabular}{lllclclclclc}
\toprule
\multirow{3}{*}{\textbf{DATASET}} & \multirow{3}{*}{\textbf{\makecell{METHOD\\ TYPE}}} & \multicolumn{2}{c}{\multirow{2}{*}{\textbf{AUROC}}}  &  \multicolumn{4}{c}{\textbf{AUPR}} & \multicolumn{4}{c}{\textbf{Pr@P}} \\ 
\cmidrule(r){5-8} \cmidrule(r){9-12}
 & & \multicolumn{2}{c}{} & \multicolumn{2}{c}{\textbf{balanced}} & \multicolumn{2}{c}{\textbf{imbalanced}} & \multicolumn{2}{c}{\textbf{balanced}} & \multicolumn{2}{c}{\textbf{imbalanced}} \\ 
 \cmidrule(r){3-4} \cmidrule(r){5-6} \cmidrule(r){7-8} \cmidrule(r){9-10} \cmidrule(r){11-12}
  & &  \textbf{Method Name} & \textbf{Value}
 & \textbf{Method Name} & \textbf{Value} & \textbf{Method Name} & \textbf{Value} & \textbf{Method Name} & \textbf{Value} & \textbf{Method Name} & \textbf{Value}  \\ 
\midrule
\multirow{3}{*}{\texttt{clg-msg-t}} & local-sim & \textit{RA} & 0.678 & \textit{RA} & 0.685 & \textit{RA} & 0.206 & \textit{RA} & 0.630 & \textit{RA} & 0.245 \\
 & global-sim & \textit{PA} & 0.751 & \textit{PA} & 0.751 & \textit{PA} & 0.267 & \textit{PA} & 0.693 & \textit{PA} & 0.300 \\
 & learning & \textit{N2V\_Avg\_LR} & 0.688 & \textit{N2V\_Avg\_LR} & 0.688 & \textit{DW\_L2\_LR} & 0.189 & \textit{N2V\_Avg\_LR} & 0.643 & \textit{N2V\_Avg\_LR} & 0.236 \\
\midrule
\multirow{3}{*}{\texttt{email-t}} & local-sim & \textit{RA} & 0.860 & \textit{RA} & 0.846 & \textit{RA} & 0.432 & \textit{RA} & 0.786 & \textit{RA} & 0.464 \\
 & global-sim & \textit{Katz} & 0.777 & \textit{Katz} & 0.753 & \textit{Katz} & 0.270 & \textit{Katz} & 0.716 & \textit{Katz} & 0.310 \\
 & learning & \textit{GS\_MaxPool\_Hada\_RF} & 0.803 & \textit{DW\_L2\_LR} & 0.764 & \textit{DW\_L2\_LR} & 0.278 & \textit{GS\_MaxPool\_Hada\_RF} & 0.742 & \textit{DW\_L1\_LR} & 0.348 \\
\midrule
\multirow{3}{*}{\texttt{math-flow-t}} & local-sim & \textit{RA} & 0.889 & \textit{RA} & 0.902 & \textit{RA} & 0.624 & \textit{RA} & 0.811 & \textit{RA} & 0.589 \\
 & global-sim & \textit{PA} & 0.912 & \textit{PA} & 0.911 & \textit{PA} & 0.621 & \textit{PA} & 0.829 & \textit{PA} & 0.579 \\
 & learning & \textit{DW\_Avg\_LR} & 0.871 & \textit{DW\_Avg\_LR} & 0.860 & \textit{DW\_Avg\_LR} & 0.438 & \textit{DW\_Avg\_LR} & 0.793 & \textit{DW\_Avg\_LR} & 0.465 \\
\midrule
\multirow{3}{*}{\texttt{ast-phys-t}} & local-sim & \textit{CN} & 0.789 & \textit{RA} & 0.716 & \textit{RA} & 0.196 & \textit{AA} & 0.721 & \textit{RA} & 0.184 \\
 & global-sim & \textit{PA} & 0.941 & \textit{PA} & 0.908 & \textit{PA} & 0.485 & \textit{PA} & 0.856 & \textit{PA} & 0.580 \\
 & learning & \textit{N2V\_Avg\_LR} & 0.634 & \textit{N2V\_Avg\_RF} & 0.974 & \textit{N2V\_Avg\_LR} & 0.756 & \textit{N2V\_Avg\_RF} & 0.945 & \textit{N2V\_Avg\_RF} & 0.736 \\
\midrule
\multirow{3}{*}{\texttt{lst-fm-t}} & local-sim & \textit{RA} & 0.591 & \textit{RA} & 0.578 & \textit{RA} & 0.126 & \textit{RA} & 0.565 & \textit{RA} & 0.151 \\
 & global-sim & \textit{PA} & 0.598 & \textit{PA} & 0.581 & \textit{PA} & 0.127 & \textit{PA} & 0.570 & \textit{PA} & 0.149 \\
 & learning & \textit{N2V\_Avg\_LR} & 0.576 & \textit{N2V\_Avg\_LR} & 0.565 & \textit{GS\_Mean\_Avg\_RF} & 0.120 & \textit{N2V\_Avg\_LR} & 0.557 & \textit{GS\_Mean\_Avg\_LR} & 0.148 \\
\midrule
\multirow{3}{*}{\texttt{iacontact-t}} & local-sim & \textit{RA} & 0.608 & \textit{RA} & 0.608 & \textit{RA} & 0.153 & \textit{AA} & 0.573 & \textit{RA} & 0.180 \\
 & global-sim & \textit{PA} & 0.583 & \textit{PA} & 0.584 & \textit{PA} & 0.133 & \textit{PA} & 0.575 & \textit{PA} & 0.167 \\
 & learning & \textit{GS\_LSTM\_L1\_RF} & 0.612 & \textit{GS\_LSTM\_L1\_RF} & 0.614 & \textit{GS\_LSTM\_L1\_RF} & 0.134 & \textit{GS\_LSTM\_L1\_RF} & 0.618 & \textit{GS\_Mean\_L2\_RF} & 0.146 \\
\midrule
\multirow{3}{*}{\texttt{forum-t}} & local-sim & \textit{AA} & 0.608 & \textit{AA} & 0.606 & \textit{AA} & 0.146 & \textit{AA} & 0.565 & \textit{AA} & 0.179 \\
 & global-sim & \textit{PA} & 0.710 & \textit{PA} & 0.673 & \textit{PA} & 0.193 & \textit{PA} & 0.642 & \textit{PA} & 0.241 \\
 & learning & \textit{DW\_Avg\_LR} & 0.641 & \textit{DW\_Avg\_RF} & 0.619 & \textit{GS\_MeanPool\_Avg\_RF} & 0.151 & \textit{DW\_Avg\_LR} & 0.596 & \textit{DW\_Avg\_RF} & 0.197 \\
\midrule
\multirow{3}{*}{\texttt{topo-t}} & local-sim & \textit{AA} & 0.938 & \textit{AA} & 0.945 & \textit{AA} & 0.609 & \textit{AA} & 0.857 & \textit{AA} & 0.581 \\
 & global-sim & \textit{PA} & 0.952 & \textit{PA} & 0.951 & \textit{PA} & 0.617 & \textit{PA} & 0.890 & \textit{PA} & 0.588 \\
 & learning & \textit{DW\_L2\_LR} & 0.914 & \textit{DW\_Avg\_LR} & 0.899 & \textit{GS\_MaxPool\_Hada\_RF} & 0.552 & \textit{DW\_Avg\_LR} & 0.832 & \textit{DW\_L1\_LR} & 0.570 \\
 \midrule
\texttt{act-mooc-t} & local-sim & \textit{RA} & 0.738 & \textit{RA} & 0.713 & \textit{RA} & 0.23 & \textit{CN} & 0.660 & \textit{RA} & 0.237 \\
 & global-sim & \textit{PA} & 0.974 & \textit{PA} & 0.963 & \textit{PA} & 0.763 & \textit{PA} & 0.945 & \textit{PA} & 0.836 \\
 & learning & \textit{DW\_Avg\_LR} & 0.977 & \textit{N2V\_Hada\_RF} & 0.976 & \textit{N2V\_Hada\_LR} & 0.857 & \textit{N2V\_Avg\_RF} & 0.945 & \textit{N2V\_L1\_LR} & 0.833 \\
\bottomrule
\end{tabular}
\end{adjustbox}
\end{table*}

\begin{landscape}

\begin{table*}[h]
\centering
\caption{Best performing methods in each group: local-sim, global-sim and learning, for missing LP in test set of two-hop away node pairs}
\begin{adjustbox}{width=1.1\textwidth,center}
\begin{tabular}{lllclclclclc}
\toprule
\multirow{3}{*}{\textbf{DATASET}} & \multirow{3}{*}{\textbf{\makecell{METHOD\\ TYPE}}} & \multicolumn{2}{c}{\multirow{2}{*}{\textbf{AUROC}}}  &  \multicolumn{4}{c}{\textbf{AUPR}} & \multicolumn{4}{c}{\textbf{Pr@P}} \\ 
\cmidrule(r){5-8} \cmidrule(r){9-12}
 & & \multicolumn{2}{c}{} & \multicolumn{2}{c}{\textbf{balanced}} & \multicolumn{2}{c}{\textbf{imbalanced}} & \multicolumn{2}{c}{\textbf{balanced}} & \multicolumn{2}{c}{\textbf{imbalanced}} \\ 
 \cmidrule(r){3-4} \cmidrule(r){5-6} \cmidrule(r){7-8} \cmidrule(r){9-10} \cmidrule(r){11-12}
& & \textbf{Method Name} & \textbf{Value} & \textbf{Method Name} & \textbf{Value} & \textbf{Method Name} & \textbf{Value} & \textbf{Method Name} & \textbf{Value} & \textbf{Method Name} & \textbf{Value} \\
\midrule
\texttt{clg-msg} & local-sim & \textit{RA} & 0.721 & \textit{RA} & 0.720 & \textit{RA} & 0.222 & \textit{RA} & 0.687 & \textit{RA} & 0.281 \\
& global-sim & \textit{Katz} & 0.799 & \textit{Katz} & 0.805 & \textit{Katz} & 0.280 & \textit{Katz} & 0.746 & \textit{Katz} & 0.356 \\
& learning & \textit{DW\_L2\_LR} & 0.661 & \textit{DW\_L2\_LR} & 0.649 & \textit{DW\_L2\_LR} & 0.171 & \textit{DW\_L2\_LR} & 0.643 & \textit{DW\_L1\_LR} & 0.201 \\
\midrule
\texttt{email} & local-sim & \textit{RA} & 0.855 & \textit{RA} & 0.842 & \textit{RA} & 0.325 & \textit{RA} & 0.807 & \textit{RA} & 0.394 \\
& global-sim & \textit{Katz} & 0.791 & \textit{Katz} & 0.775 & \textit{Katz} & 0.257 & \textit{Katz} & 0.744 & \textit{Katz} & 0.322 \\
& learning & \textit{DW\_L2\_LR} & 0.776 & \textit{DW\_L2\_LR} & 0.759 & \textit{DW\_L2\_LR} & 0.231 & \textit{DW\_L2\_LR} & 0.715 & \textit{DW\_L2\_LR} & 0.278 \\
\midrule
\texttt{math-flow} & local-sim & \textit{AA} & 0.878 & \textit{AA} & 0.885 & \textit{AA} & 0.545 & \textit{AA} & 0.808 & \textit{AA} & 0.546 \\
& global-sim & \textit{PA} & 0.884 & \textit{PA} & 0.879 & \textit{PA} & 0.511 & \textit{KPA} & 0.800 & \textit{Katz} & 0.529 \\
& learning & \textit{DW\_L2\_LR} & 0.953 & \textit{DW\_L2\_LR} & 0.809 & \textit{DW\_L2\_LR} & 0.380 & \textit{DW\_L2\_LR} & 0.729 & \textit{DW\_L2\_LR} & 0.437 \\
\midrule
\texttt{ast-phys} & local-sim & \textit{RA} & 0.920 & \textit{RA} & 0.952 & \textit{RA} & 0.661 & \textit{RA} & 0.900 & \textit{RA} & 0.631 \\
& global-sim & \textit{PA} & 0.984 & \textit{CT} & 0.828 & \textit{PA} & 0.826 & \textit{CT} & 0.793 & \textit{PA} & 0.820 \\
& learning & \textit{N2V\_L2\_LR} & 0.953 & \textit{GS\_Mean\_L1\_LR} & 0.819 & \textit{N2V\_L1\_LR} & 0.620 & \textit{GS\_MeanPool\_L2\_LR} & 0.745 & \textit{N2V\_L2\_LR} & 0.650 \\
\midrule
\texttt{lst-fm} & local-sim & \textit{RA} & 0.667 & \textit{RA} & 0.621 & \textit{AA} & 0.147 & \textit{RA} & 0.617 & \textit{CN} & 0.178 \\
& global-sim & \textit{PA} & 0.675 & \textit{PA} & 0.626 & \textit{PA} & 0.150 & \textit{PA} & 0.622 & \textit{PA} & 0.177 \\
& learning & \textit{DW\_L2\_LR} & 0.623 & \textit{DW\_L2\_LR} & 0.577 & \textit{DW\_L2\_LR} & 0.125 & \textit{DW\_L2\_LR} & 0.585 & \textit{DW\_L1\_RF} & 0.147 \\
\midrule
\texttt{iacontact} & local-sim & \textit{RA} & 0.608 & \textit{AA} & 0.516 & \textit{JC} & 0.118 & \textit{CN} & 0.505 & \textit{CN} & 0.123 \\
& global-sim & \textit{PA} & 0.592 & \textit{PA} & 0.517 & \textit{PA} & 0.112 & \textit{PA} & 0.505 & \textit{PA} & 0.112 \\
& learning & \textit{GS\_LSTM\_Avg\_LR} & 0.560 & \textit{GS\_LSTM\_L1\_LR} & 0.573 & \textit{GS\_MeanPool\_Hada\_LR} & 0.103 & \textit{GS\_LSTM\_L1\_RF} & 0.595 & \textit{GS\_MaxPool\_Avg\_RF} & 0.123 \\
\midrule
\texttt{forum} & local-sim & \textit{AA} & 0.617 & \textit{RA} & 0.629 & \textit{CN} & 0.167 & \textit{RA} & 0.590 & \textit{CN} & 0.198 \\
& global-sim & \textit{PA} & 0.732 & \textit{PA} & 0.698 & \textit{PA} & 0.323 & \textit{PA} & 0.658 & \textit{PA} & 0.260 \\
& learning & \textit{DW\_L2\_LR} & 0.588 & \textit{DW\_Hada\_RF} & 0.580 & \textit{DW\_L2\_RF} & 0.137 & \textit{DW\_L2\_LR} & 0.602 & \textit{DW\_L1\_RF} & 0.167 \\
\midrule
\texttt{topo} & local-sim & \textit{RA} & 0.848 & \textit{RA} & 0.897 & \textit{RA} & 0.603 & \textit{RA} & 0.811 & \textit{RA} & 0.614 \\
& global-sim & \textit{PA} & 0.922 & \textit{PA} & 0.911 & \textit{PA} & 0.608 & \textit{PA} & 0.831 & \textit{PA} & 0.697 \\
& learning & \textit{GS\_MaxPool\_Hada\_RF} & 0.818 & \textit{GS\_MaxPool\_Hada\_RF} & 0.804 & \textit{DW\_Avg\_RF} & 0.410 & \textit{GS\_MaxPool\_Hada\_RF} & 0.743 & \textit{DW\_Avg\_RF} & 0.408 \\
\midrule
\texttt{act-mooc} & local-sim & \textit{RA} & 0.840 & \textit{RA} & 0.834 & \textit{RA} & 0.398 & \textit{RA} & 0.771 & \textit{AA} & 0.378 \\
& global-sim & \textit{PA} & 0.994 & \textit{PA} & 0.991 & \textit{PA} & 0.915 & \textit{PA} & 0.984 & \textit{PA} & 0.894 \\
& learning & \textit{DW\_Avg\_LR} & 0.989 & \textit{DW\_Avg\_LR} & 0.984 & \textit{DW\_Avg\_LR} & 0.840 & \textit{N2V\_L1\_LR} & 0.975 & \textit{DW\_Avg\_LR} & 0.829 \\
\midrule
\texttt{arxiv} & local-sim & \textit{RA} & 0.951 & \textit{RA} & 0.952 & \textit{RA} & 0.662 & \textit{RA} & 0.900 & \textit{RA} & 0.677 \\
& global-sim & \textit{Norm-CT} & 0.854 & \textit{CT} & 0.828 & \textit{CT} & 0.339 & \textit{CT} & 0.793 & \textit{Katz} & 0.427 \\
& learning & \textit{N2V\_L2\_LR} & 0.797 & \textit{GS\_Mean\_L1\_LR} & 0.819 & \textit{DW\_L1\_LR} & 0.376 & \textit{GS\_MeanPool\_L2\_LR} & 0.745 & \textit{DW\_L1\_LR} & 0.421 \\
\midrule
\texttt{pow-grid} & local-sim & \textit{CN} & 0.570 & \textit{CN} & 0.772 & \textit{CN} & 0.314 & \textit{AA} & 0.574 & \textit{CN} & 0.179 \\
& global-sim & \textit{Katz} & 0.606 & \textit{Katz} & 0.682 & \textit{Katz} & 0.195 & \textit{PA} & 0.617 & \textit{Katz} & 0.265 \\
& learning & \textit{GS\_MaxPool\_L1\_LR} & 0.616 & \textit{GS\_MaxPool\_L2\_LR} & 0.693 & \textit{GS\_MaxPool\_L1\_LR} & 0.157 & \textit{GS\_Mean\_L1\_LR} & 0.679 & \textit{GS\_MaxPool\_L2\_RF} & 0.240 \\
\midrule
\texttt{routers} & local-sim & \textit{AA} & 0.833 & \textit{AA} & 0.871 & \textit{AA} & 0.431 & \textit{AA} & 0.803 & \textit{AA} & 0.452 \\
& global-sim & \textit{Katz} & 0.760 & \textit{Katz} & 0.834 & \textit{Katz} & 0.438 & \textit{Katz} & 0.785 & \textit{Katz} & 0.433 \\
& learning & \textit{GS\_LSTM\_L1\_LR} & 0.690 & \textit{DW\_L2\_RF} & 0.738 & \textit{DW\_L2\_LR} & 0.204 & \textit{GS\_Mean\_L2\_LR} & 0.734 & \textit{DW\_L1\_RF} & 0.259 \\
\midrule
\texttt{bio-yeast} & local-sim & \textit{RA} & 0.877 & \textit{RA} & 0.882 & \textit{RA} & 0.525 & \textit{AA} & 0.805 & \textit{RA} & 0.535 \\
& global-sim & \textit{Katz} & 0.863 & \textit{Katz} & 0.860 & \textit{Katz} & 0.465 & \textit{Katz} & 0.781 & \textit{Katz} & 0.471 \\
& learning & \textit{DW\_L2\_LR} & 0.852 & \textit{DW\_L2\_LR} & 0.863 & \textit{DW\_L2\_LR} & 0.466 & \textit{DW\_L2\_LR} & 0.765 & \textit{DW\_L2\_LR} & 0.467 \\
\midrule
\texttt{fb} & local-sim & \textit{RA} & 0.945 & \textit{RA} & 0.956 & \textit{RA} & 0.630 & \textit{RA} & 0.925 & \textit{RA} & 0.636 \\
& global-sim & \textit{Norm-CT} & 0.903 & \textit{Katz} & 0.919 & \textit{Katz} & 0.528 & \textit{Katz} & 0.886 & \textit{Katz} & 0.543 \\
& learning & \textit{DW\_L2\_LR} & 0.914 & \textit{DW\_L1\_RF} & 0.928 & \textit{DW\_L1\_LR} & 0.554 & \textit{DW\_L2\_LR} & 0.888 & \textit{DW\_L2\_LR} & 0.582 \\
\midrule
\texttt{blog-cat} & local-sim & \textit{RA} & 0.932 & \textit{RA} & 0.935 & \textit{RA} & 0.675 & \textit{RA} & 0.872 & \textit{RA} & 0.656 \\
& global-sim & \textit{PA} & 0.921 & \textit{PA} & 0.925 & \textit{PA} & 0.642 & \textit{PA} & 0.885 & \textit{PA} & 0.625 \\
& learning & \textit{DW\_L1\_LR} & 0.949 & \textit{DW\_Avg\_LR} & 0.903 & \textit{DW\_L1\_LR} & 0.723 & \textit{DW\_Avg\_LR} & 0.834 & \textit{DW\_Avg\_LR} & 0.581 \\
\bottomrule
\end{tabular}
\end{adjustbox}

\label{tab:results}
\end{table*}
\end{landscape}

\begin{table*}[h]
\caption{Best performing methods in each group: global-sim and learning, for future LP in test set of three-hop away node pairs}
\renewcommand{\arraystretch}{2.5} 
\begin{adjustbox}{width=1\textwidth,center}
\begin{tabular}{l l l c l c l c l c l c}
\toprule
\multirow{3}{*}{\textbf{DATASET}} & \multirow{3}{*}{\textbf{\makecell{METHOD\\ TYPE}}} & \multicolumn{2}{c}{\multirow{2}{*}{\textbf{AUROC}}}  &  \multicolumn{4}{c}{\textbf{AUPR}} & \multicolumn{4}{c}{\textbf{Pr@P}} \\ 
\cmidrule(r){5-8} \cmidrule(r){9-12}
 & & \multicolumn{2}{c}{} & \multicolumn{2}{c}{\textbf{balanced}} & \multicolumn{2}{c}{\textbf{imbalanced}} & \multicolumn{2}{c}{\textbf{balanced}} & \multicolumn{2}{c}{\textbf{imbalanced}} \\ 
 \cmidrule(r){3-4} \cmidrule(r){5-6} \cmidrule(r){7-8} \cmidrule(r){9-10} \cmidrule(r){11-12}
& & \textbf{Method Name} & \textbf{Value} & \textbf{Method Name} & \textbf{Value} & \textbf{Method Name} & \textbf{Value} & \textbf{Method Name} & \textbf{Value} & \textbf{Method Name} & \textbf{Value} \\
\midrule
\texttt{clg-msg-t} & global-sim & \textit{PA} & 0.753 & \textit{PA} & 0.737 & \textit{PA} & 0.268 & \textit{PA} & 0.685 & \textit{Katz} & 0.316 \\
 & learning & \textit{GS\_MeanPool\_Avg\_RF} & 0.696 & \textit{DW\_Avg\_LR} & 0.688 & \textit{GS\_MeanPool\_Avg\_RF} & 0.187 & \textit{GS\_MeanPool\_Avg\_RF} & 0.648 & \textit{DW\_Avg\_RF} & 0.243 \\
\midrule
\texttt{email-t} & global-sim & \textit{Katz} & 0.742 & \textit{Katz} & 0.720 & \textit{Katz} & 0.218 & \textit{Katz} & 0.658 & \textit{Katz} & 0.286 \\
 & learning & \textit{GS\_Mean\_L1\_RF} & 0.793 & \textit{GS\_Mean\_L1\_RF} & 0.794 & \textit{GS\_MeanPool\_Hada\_RF} & 0.326 & \textit{GS\_Mean\_L1\_LR} & 0.748 & \textit{GS\_LSTM\_L1\_LR} & 0.396 \\
\midrule
\texttt{math-flow-t} & global-sim & \textit{PA} & 0.888 & \textit{PA} & 0.883 & \textit{PA} & 0.531 & \textit{PA} & 0.806 & \textit{PA} & 0.525 \\
 & learning & \textit{N2V\_Avg\_LR} & 0.812 & \textit{DW\_Avg\_LR} & 0.811 & \textit{Dw\_Avg\_LR} & 0.381 & \textit{DW\_Avg\_LR} & 0.742 & \textit{DW\_Avg\_LR} & 0.408 \\
\midrule
\texttt{ast-phys-t} & global-sim & \textit{Katz} & 0.941 & \textit{Katz} & 0.944 & \textit{Katz} & 0.754 & \textit{Katz} & 0.973 & \textit{PA} & 0.684 \\
 & learning & \textit{N2V\_Avg\_RF} & 0.987 & \textit{N2V\_Avg\_RF} & 0.983 & \textit{N2V\_Avg\_RF} & 0.919 & \textit{N2V\_L2\_LR} & 0.990 & \textit{N2V\_Avg\_RF} & 0.833 \\
\midrule
\texttt{lst-fm-t} & global-sim & \textit{CT} & 0.504 & \textit{PA} & 0.733 & \textit{PA} & 0.195 & \textit{PA} & 0.913 & \textit{Norm-CT} & 0.168 \\
 & learning & \textit{GS\_Mean\_Avg\_RF} & 0.734 & \textit{GS\_Mean\_L2\_LR} & 0.723 & \textit{GS\_MaxPool\_Hada\_RF} & 0.264 & \textit{GS\_Mean\_Hada\_RF} & 0.940 & \textit{GS\_Mean\_L2\_LR} & 0.309 \\
\midrule
\texttt{forum-t} & global-sim & \textit{PA} & 0.706 & \textit{PA} & 0.672 & \textit{PA} & 0.185 & \textit{PA} & 0.638 & \textit{PA} & 0.227 \\
 & learning & \textit{DW\_Avg\_LR} & 0.649 & \textit{DW\_L2\_LR} & 0.629 & \textit{DW\_L1\_LR} & 0.153 & \textit{DW\_Avg\_LR} & 0.610 & \textit{DW\_L1\_LR} & 0.175 \\
\midrule
\texttt{topo-t} & global-sim & \textit{PA} & 0.905 & \textit{PA} & 0.906 & \textit{PA} & 0.550 & \textit{PA} & 0.837 & \textit{PA} & 0.533 \\
 & learning & \textit{GS\_MeanPool\_Avg\_RF} & 0.906 & \textit{GS\_MaxPool\_Avg\_RF} & 0.903 & \textit{GS\_MeanPool\_Hada\_RF} & 0.609 & \textit{GS\_MaxPool\_Avg\_RF} & 0.827 & \textit{GS\_MaxPool\_Avg\_RF} & 0.609 \\
\midrule
\texttt{act-mooc-t} & global-sim & \textit{PA} & 0.967 & \textit{PA} & 0.983 & \textit{PA} & 0.910 & \textit{PA} & 0.969 & \textit{PA} & 0.893 \\
 & learning & \textit{N2V\_L1\_LR} & 0.964 & \textit{N2V\_L1\_LR} & 0.966 & \textit{N2V\_L1\_LR} & 0.902 & \textit{N2V\_L1\_LR} & 0.969 & \textit{N2V\_L1\_LR} & 0.893 \\
\hline
\end{tabular}
\end{adjustbox}
\end{table*}

\begin{table*}[h]
\caption{Best performing methods in each group: global-sim and learning, for missing LP in test set of three-hop away node pairs}
\renewcommand{\arraystretch}{2.2} 
\begin{adjustbox}{width=1\textwidth}
\begin{tabular}{l l l c l c l c l c l c}
\toprule
\multirow{3}{*}{\textbf{DATASET}} & \multirow{3}{*}{\textbf{\makecell{METHOD\\ TYPE}}} & \multicolumn{2}{c}{\multirow{2}{*}{\textbf{AUROC}}}  &  \multicolumn{4}{c}{\textbf{AUPR}} & \multicolumn{4}{c}{\textbf{Pr@P}} \\ 
\cmidrule(r){5-8} \cmidrule(r){9-12}
 & & \multicolumn{2}{c}{} & \multicolumn{2}{c}{\textbf{balanced}} & \multicolumn{2}{c}{\textbf{imbalanced}} & \multicolumn{2}{c}{\textbf{balanced}} & \multicolumn{2}{c}{\textbf{imbalanced}} \\ 
 \cmidrule(r){3-4} \cmidrule(r){5-6} \cmidrule(r){7-8} \cmidrule(r){9-10} \cmidrule(r){11-12}
& & \textbf{Method Name} & \textbf{Value} & \textbf{Method Name} & \textbf{Value} & \textbf{Method Name} & \textbf{Value} & \textbf{Method Name} & \textbf{Value} & \textbf{Method Name} & \textbf{Value} \\
\midrule
\texttt{clg-msg} & global-sim & \textit{PA} & 0.887 & \textit{PA} & 0.887 & \textit{PA} & 0.527 & \textit{PA} & 0.823 & \textit{PA} & 0.535 \\
                 & learning   & \textit{DW\_L2\_LR} & 0.648 & \textit{DW\_L1\_RF} & 0.669 & \textit{DW\_L1\_RF} & 0.178 & \textit{DW\_L2\_LR} & 0.640 & \textit{DW\_L1\_RF} & 0.235 \\
\midrule
\texttt{email}   & global-sim & \textit{Katz} & 0.611 & \textit{Norm-CT} & 0.775 & \textit{Katz} & 0.151 & \textit{Katz} & 0.583 & \textit{Katz} & 0.250 \\
                 & learning   & \textit{GS\_Attentional\_Avg\_LR} & 0.675 & \textit{GS\_MaxPool\_L1\_LR} & 0.722 & \textit{GS\_LSTM\_L2\_RF} & 0.252 & \textit{N2V\_Hada\_RF} & 0.666 & \textit{GS\_Mean\_L1\_RF} & 0.333 \\
\midrule
\texttt{math-flow} & global-sim & \textit{PA} & 0.886 & \textit{PA} & 0.869 & \textit{PA} & 0.489 & \textit{PA} & 0.952 & \textit{PA} & 0.476 \\
                   & learning   & \textit{GS\_MeanPool\_Hada\_LR} & 0.719 & \textit{GS\_MeanPool\_Hada\_LR} & 0.714 & \textit{GS\_MeanPool\_L1\_LR} & 0.224 & \textit{GS\_MaxPool\_L2\_LR} & 0.928 & \textit{GS\_MaxPool\_L1\_LR} & 0.291 \\
\midrule
\texttt{ast-phys} & global-sim & \textit{PA} & 0.991 & \textit{PA} & 0.996 & \textit{Katz} & 0.906 & \textit{PA} & 0.971 & \textit{PA} & 0.857 \\
                  & learning   & \textit{N2V\_L2\_RF} & 0.924 & \textit{GS\_MaxPool\_L1\_RF} & 0.787 & \textit{N2V\_L2\_RF} & 0.584 & \textit{N2V\_L1\_LR} & 0.971 & \textit{N2V\_L2\_RF} & 0.571 \\
\midrule
\texttt{forum}    & global-sim & \textit{Katz} & 0.742 & \textit{Katz} & 0.779 & \textit{Katz} & 0.250 & \textit{Katz} & 0.682 & \textit{Katz} & 0.346 \\
                  & learning   & \textit{DW\_L2\_LR} & 0.625 & \textit{GS\_MaxPool\_Hada\_LR} & 0.601 & \textit{DW\_L2\_LR} & 0.127 & \textit{DW\_L2\_LR} & 0.567 & \textit{DW\_L1\_RF} & 0.173 \\
\midrule
\texttt{topo}     & global-sim & \textit{PA} & 0.879 & \textit{PA} & 0.873 & \textit{CT} & 0.492 & \textit{PA} & 0.779 & \textit{Norm-CT} & 0.485 \\
                  & learning   & \textit{GS\_MeanPool\_Hada\_RF} & 0.448 & \textit{GS\_Mean\_Hada\_RF} & 0.880 & \textit{GS\_MeanPool\_Hada\_RF} & 0.549 & \textit{GS\_MeanPool\_Hada\_RF} & 0.805 & \textit{GS\_Mean\_Hada\_RF} & 0.552 \\
\midrule
\texttt{pow-grid} & global-sim & \textit{Norm-CT} & 0.616 & \textit{CT} & 0.594 & \textit{CT} & 0.119 & \textit{Norm-CT} & 0.602 & \textit{-Norm-CT} & 0.136 \\
                  & learning   & \textit{GS\_Mean\_L2\_RF} & 0.596 & \textit{GS\_Mean\_L2\_RF} & 0.723 & \textit{GS\_MaxPool\_L1\_LR} & 0.126 & \textit{GS\_Mean\_L1\_RF} & 0.630 & \textit{N2V\_L1\_RF} & 0.150 \\
\midrule
\texttt{routers}  & global-sim & \textit{Norm-CT} & 0.748 & \textit{Norm-CT} & 0.809 & \textit{Norm-CT} & 0.288 & \textit{CT} & 0.716 & \textit{CT} & 0.333 \\
                  & learning   & \textit{GS\_MaxPool\_Hada\_LR} & 0.720 & \textit{DW\_Hada\_LR} & 0.711 & \textit{GS\_MaxPool\_L1\_LR} & 0.216 & \textit{GS\_LSTM\_L2\_LR} & 0.641 & \textit{GS\_MaxPool\_L1\_LR} & 0.320 \\
\midrule
\texttt{boi-yeast} & global-sim & \textit{PA} & 0.762 & \textit{PA} & 0.715 & \textit{CT} & 0.244 & \textit{PA} & 0.933 & \textit{Norm-CT} & 0.311 \\
                   & learning   & \textit{DW\_L2\_LR} & 0.721 & \textit{DW\_L1\_LR} & 0.696 & \textit{GS\_LSTM\_Hada\_RF} & 0.231 & \textit{DW\_L2\_LR} & 0.933 & \textit{GS\_Mean\_Avg\_RF} & 0.255 \\
\midrule
\texttt{fb}       & global-sim & \textit{Norm-CT} & 0.942 & \textit{Katz} & 0.944 & \textit{-NormCT} & 0.700 & \textit{PA} & 0.875 & \textit{Katz} & 0.250 \\
                  & learning   & \textit{N2V\_L1\_LR} & 0.948 & \textit{GS\_MaxPool\_L2\_RF} & 0.935 & \textit{GS\_MaxPool\_Hada\_RF} & 0.713 & \textit{DW\_Hada\_LR} & 0.100 & \textit{DW\_Hada\_LR} & 0.625 \\
\midrule
\texttt{blog-cat} & global-sim & \textit{PA} & 0.807 & \textit{PA} & 0.801 & \textit{Norm-CT} & 0.392 & \textit{PA} & 0.947 & \textit{PA} & 0.391 \\
                  & learning   & \textit{N2V\_Avg\_RF} & 0.792 & \textit{N2V\_Avg\_LR} & 0.742 & \textit{N2V\_Avg\_RF} & 0.310 & \textit{N2V\_Avg\_RF} & 0.939 & \textit{N2V\_Avg\_RF} & 0.347 \\
\bottomrule
\end{tabular}
\end{adjustbox}
\end{table*}

\end{appendices}
%
%

\end{document}